\documentclass[twocolumn]{aastex62}
\pdfoutput=1
\usepackage[utf8]{inputenc}
\usepackage{amsmath}
\usepackage{mathtools}
\usepackage{graphicx}
\usepackage{bm}
\usepackage{tabularx}

\newcommand{\disphat}[2][4mu]{\hat{#2\mkern#1}\mkern-#1}

\makeatletter
\newcommand{\pushright}[1]{\ifmeasuring@#1\else\omit\hfill$\displaystyle#1$\fi\ignorespaces}
\makeatother

\newcommand{\evec}{\bm{{e}}}
\newcommand{\ehat}{\bm{{\hat{e}}}}
\newcommand{\jvec}{\bm{{j}}}
\newcommand{\jhat}{\bm{{\disphat{j}}}}

\newcommand{\qvec}{\bm{{q}}}
\newcommand{\qhat}{\bm{{\hat{q}}}}
\newcommand{\nhat}{\bm{{\hat{n}}}}
\newcommand{\rvec}{\bm{{r}}}
\newcommand{\rhat}{\bm{{\hat{r}}}}

\newcommand{\sortletter}[1]{}

\submitjournal{\apj}

\shorttitle{Compact-Object Mergers in the Galactic Center}
\shortauthors{Bub \& Petrovich}

\begin{document}

\title{Compact-Object Mergers in the Galactic Center: Evolution in Triaxial Clusters}

\correspondingauthor{Mathew Bub}
\email{mathew.bub@mail.utoronto.ca}

\author[0000-0003-4251-0845]{Mathew W. Bub}
\affiliation{Canadian Institute for Theoretical Astrophysics, University of Toronto, 60 St. George Street, Toronto, ON M5S 3H8, Canada}

\author[0000-0003-0412-9314]{Cristobal Petrovich}
\affiliation{Canadian Institute for Theoretical Astrophysics, University of Toronto, 60 St. George Street, Toronto, ON M5S 3H8, Canada}
\affiliation{Centre for Planetary Sciences, Department of Physical \& Environmental Sciences, University of Toronto Scarborough, Toronto, ON M1C 1A4, Canada}
\affiliation{Steward Observatory, University of Arizona, 933 N. Cherry Ave., Tucson, AZ 85721, USA}

\begin{abstract}
There is significant observational evidence that a large fraction of galactic centers, including those in the Milky Way and M31, host a supermassive black hole (SMBH) embedded in a triaxial nuclear star cluster. In this work, we study the secular orbital evolution of binaries in these environments, and characterize the regions and morphological properties of nuclear star clusters that lead to gravitational wave mergers and/or tidal captures. We show that even a modest level of triaxiality in the density distribution of a cluster (an ellipsoid with axis ratios of $0.7$ and $0.95$) dramatically enhances the merger rates in the central parsecs of the Galaxy by a factor of up to $\sim10-30$ relative to a spherical density distribution. Moreover, we show that the merger fraction of binaries with semi-major axes in the range 10-100 AU remains above 10\% for the entire central parsec of the cluster, reaching values close to unity at a distance of $\sim0.2-0.4$ pc from the SMBH. We understand this large merger efficiency in terms of two distinct mechanisms: i) eccentricity oscillations driven by the dominant axisymmetric part of the cluster potential that are enhanced by the slow modulation of a binary's angular momentum from the triaxial contribution, similar to the well-known octupole-level dynamics in three-body systems; ii) chaotic diffusion of eccentricities arising when the nodal precession timescale of a binary's orbit about the SMBH becomes comparable to its  characteristic secular timescale. Overall, our results indicate that galactic centers are significantly more collisional than previously thought, with mergers taking place up to the effective radii of their nuclear star clusters.
\end{abstract}


\section{Introduction} \label{sec:intro}
\subsection{Nuclear Star Clusters}

Most nearby galaxies contain a dense stellar cluster at their kinematical and photometric centers \citep[e.g.,][]{Neumayer2011,Turner2012,Georgiev2014}. These so-called nuclear star clusters have masses in the range $10^5-10^8$ solar masses ($M_\odot$), effective radii of a few parsecs, and often host a central supermassive black hole (SMBH) \citep[e.g.,][]{Georgiev2016}. This is the case in our own galaxy, which hosts a nuclear cluster and an SMBH with masses of $\sim3\times10^7\,M_\odot$ and $\sim 4\times10^6\,M_\odot$, respectively \citep{Ghez2008,Gillessen2009}. 

Nuclear star clusters in nearby galaxies are often observed to have non-spherical mass distributions \citep[e.g.,][]{Georgiev2014}. This is the case for the cluster in the Milky Way, where the diffuse light follows an ellipsoidal distribution in the central parsecs, with major axis on the plane of the Galaxy and a mean axis ratio of $0.7-0.8$ \citep{schodel2014,fritz2016}. Furthermore, the stellar kinematics are consistent with these results \citep{chatzo2015} and additionally show evidence for triaxiality \citep{Feldmeier-Krause2017}. A more dramatic example of triaxiality is the nuclear star cluster in the Andromeda galaxy, which possesses a double nucleus \citep{Kormendy1999} that is best explained as an eccentric and apsidally aligned disk of stars \citep{tremaine1995}. In fact, these strongly triaxial nuclear structures may not be uncommon in early-type galaxies \citep{lauer2005}.

\subsection{Dynamics of Binaries and Astrophysical Applications}
The dynamics of stars in the inner parsecs of the nuclear cluster are characterized by high dispersion velocities, high stellar densities, and short relaxation timescales\footnote{The two-body relaxation timescales can be long or comparable to the age of the clusters, thus explaining why the clusters are generally not spherical. However, the resonant relaxation timescales are generally much shorter than the ages of the clusters.} \citep[e.g.,][]{merritt_book}. The dynamical evolution of binaries in these environments is a complex and multi-scale process, ranging from impulsive-like close encounters with other stars to long-range tidal torques arising from the SMBH and the cluster. Despite this complexity, there  has been significant recent progress in this field, mainly driven by the exciting possibility that a significant fraction of compact-object mergers detected by the LIGO-Virgo collaboration may arise dynamically in these extreme environments \citep[e.g.,][]{antonini2012, VanLandingham2016,bartos2017, petrovich2017, stone2017, hamers2018, hoang2018, randall2018, HR2019c, Fragione_grishin2019, zhang2019, leigh2018}. 

Beyond gravitational wave sources, the evolution of binaries in the Galactic center is likely tied to other astrophysical phenomena and stellar populations, including X-ray binaries, hypervelocity stars, S-stars, G-2 objects, and various types of transient events \citep{hills1998, hopman09, antonini2010, prodan2015, stephan2016, stephan2019, fragione_antonini2019}. The population of X-ray binaries are of particular interest to understanding the role of galactic nuclei at sourcing LIGO-Virgo events, as they serve as a proxy for the distribution of black holes in the innermost parsecs of the Galaxy. Interestingly, recent observations by \citet{hailey2018} show that a dozen detected X-ray binaries form a cusp concentrated in the central parsec of the Galaxy, implying a significant over-density of these objects in this region. This result demands an efficient formation channel that is unique to the Galactic center. One promising possibility proposed by \citet{Generozov2018} is that the X-ray binaries result from low-mass stars being tidally captured by a black hole. The authors disregard the possibility that the tidal captures (or, more precisely, dynamical hardening) can occur in binary systems, because previous works relying on the Lidov-Kozai mechanism have low capture efficiencies ($\sim1\%$) at the location of observed X-ray binaries ($\lesssim1 \, \mathrm{pc}$) \citep{prodan2015, stephan2016}. However, these works and most previous treatments for the case of gravitational wave mergers ignore the effect of the cluster potential which, as we will show, dramatically increases merger rates in this region.

\subsection{Our Work}

In this work, we study the role of the cluster potential on the secular orbital evolution of binaries. Our goal is to characterize the regions within the cluster as well as their morphological properties that lead to gravitational wave mergers and/or tidal captures.

Only recently, a few works have studied the effects of the cluster potential on the secular evolution of binaries that can lead to mergers. First, \citet{petrovich2017} considered the effect of axisymmetric clusters, and showed that mergers are greatly enhanced due to the emergence of secular chaos. However, the authors treated the cluster potential as a small perturbation to a three-body system and did not explore the full extent of the cluster, but rather limited themselves to the central $\sim0.5$ pc of a single ad-hoc potential. Second, in a series of papers \citet{HR2019a,HR2019b,HR2019c} explored the secular dynamics for the full extent of the cluster, and overcame the technical limitations of \citet{petrovich2017} by fully accounting for the cluster tidal field in the binary's evolution. However, the authors restricted themselves to non-triaxial clusters without an SMBH. We go beyond these works by considering binary evolution in the full extent of triaxial clusters with a central SMBH.



This paper is organized as follows. In Section \ref{sec:model}, we describe our model and methods. In Sections \ref{sec:torus-filling}, \ref{sec:non-torus-filling}, and \ref{sec:triaxial}, we describe our results for three distinct dynamical regimes. In Section \ref{sec:pop_synth}, we demonstrate the merger fractions in our model resulting from a population synthesis of binaries. Finally, in Section \ref{sec:discussion}, we summarize and discuss all of our main results.

\begin{figure*}[ht]
    \centering
    \includegraphics[width=\linewidth]{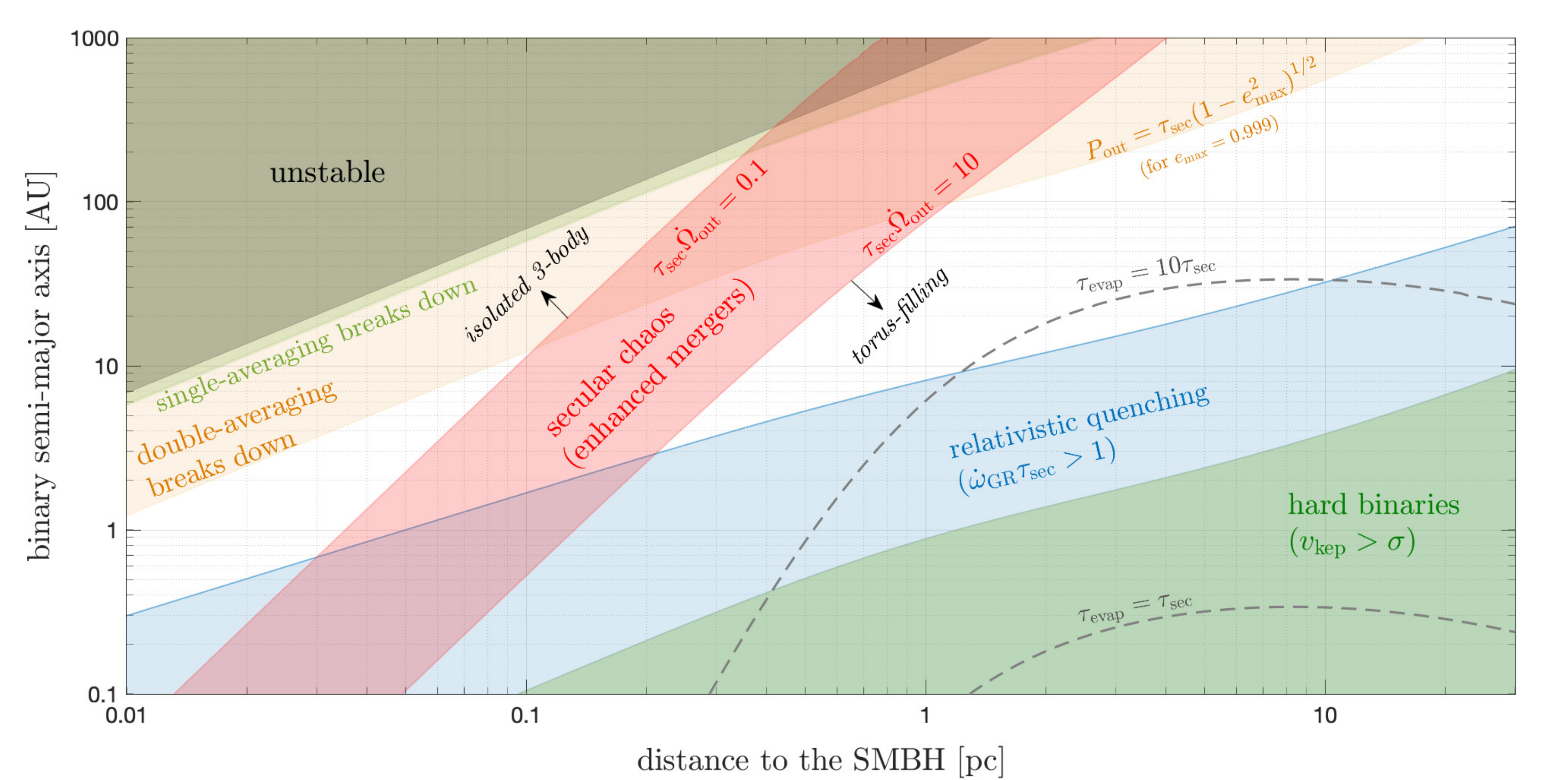}
    \caption{Dynamical regimes for a binary with total mass $M_{\rm bin}=10 \, M_\odot$ in the Milky Way Galactic center as a function of the binary semi-major axis $a_{\rm in}$ and the distance to the central black hole $a_{\rm out}$. The central black hole has a mass of $M_{\rm BH}=4\times10^{6}\,M_\odot$, while the stellar cluster is assumed to follow an axisymmetric Hernquist density profile with distance scale $s=4$ pc, total mass $8\,M_{\rm BH}$, and axis ratio $c=0.85$. From the upper-left corner to lower-right corner, the dynamical regimes are defined as follows: {\bf Unstable:} $a_{\rm in}>0.25 \, a_{\rm out}(M_{\rm bin}/M_{\rm BH})^{1/3}$ \citep{grishin2017}. {\bf Single-averaging breaks down:} $P_{\rm in}<\tau_{\rm sec}(1-e_{\rm max}^2)^{1/2}$, where averaging the forces from the SMBH and cluster over the binary's inner orbit becomes an invalid approximation (we use $e_{\rm max}=0.999$ for reference). {\bf Double-averaging breaks down:}  $P_{\rm out}<\tau_{\rm sec}(1-e_{\rm max}^2)^{1/2}$, where averaging over the outer orbit breaks down. {\bf Secular chaos:} the nodal precession of the outer orbit due to the cluster is comparable to the secular timescale, $\dot{\Omega}_{\rm out}\tau_{\rm sec}\sim0.1-10$, such that the binary evolution is neither approximated by an {\it isolated three-body system}, nor by the tidal field from a dense {\it axisymmetric torus} (Section \ref{sec:torus-filling}). This region gives rise to chaos and extreme eccentricities (Section \ref{sec:non-torus-filling}). {\bf Relativistic quenching:} $\dot\omega_{\rm GR}\tau_{\rm sec}>1$, such that the eccentricity growth from secular interactions is quenched by relativistic precession. {\bf Hard binaries:} $v_k>\sigma$, where the binaries are tight enough that they are not expected to evaporate after repeated encounters with other stars in the cluster. The dashed lines indicate the typical timescale for soft binaries to evaporate, which is longer than $\sim 10 \, \tau_{\rm sec}$ for the relevant cases of our work (i.e., not quenched by relativistic precession).}
    \label{fig:regimes}
\end{figure*}

In order to facilitate the navigation of this paper, which involves various dynamical regimes for a binary, we have provided a schematic diagram of the most relevant regimes in Figure \ref{fig:regimes} as a function of distance to the central SMBH and the binary's semi-major axis. The caption for the figure explains the different regimes, which we will discuss in detail throughout this paper.

\section{The Model} \label{sec:model}

\subsection{Coordinate System} \label{sec:coordinates}
\begin{figure}[ht]
    \centering
    \includegraphics[width=\linewidth]{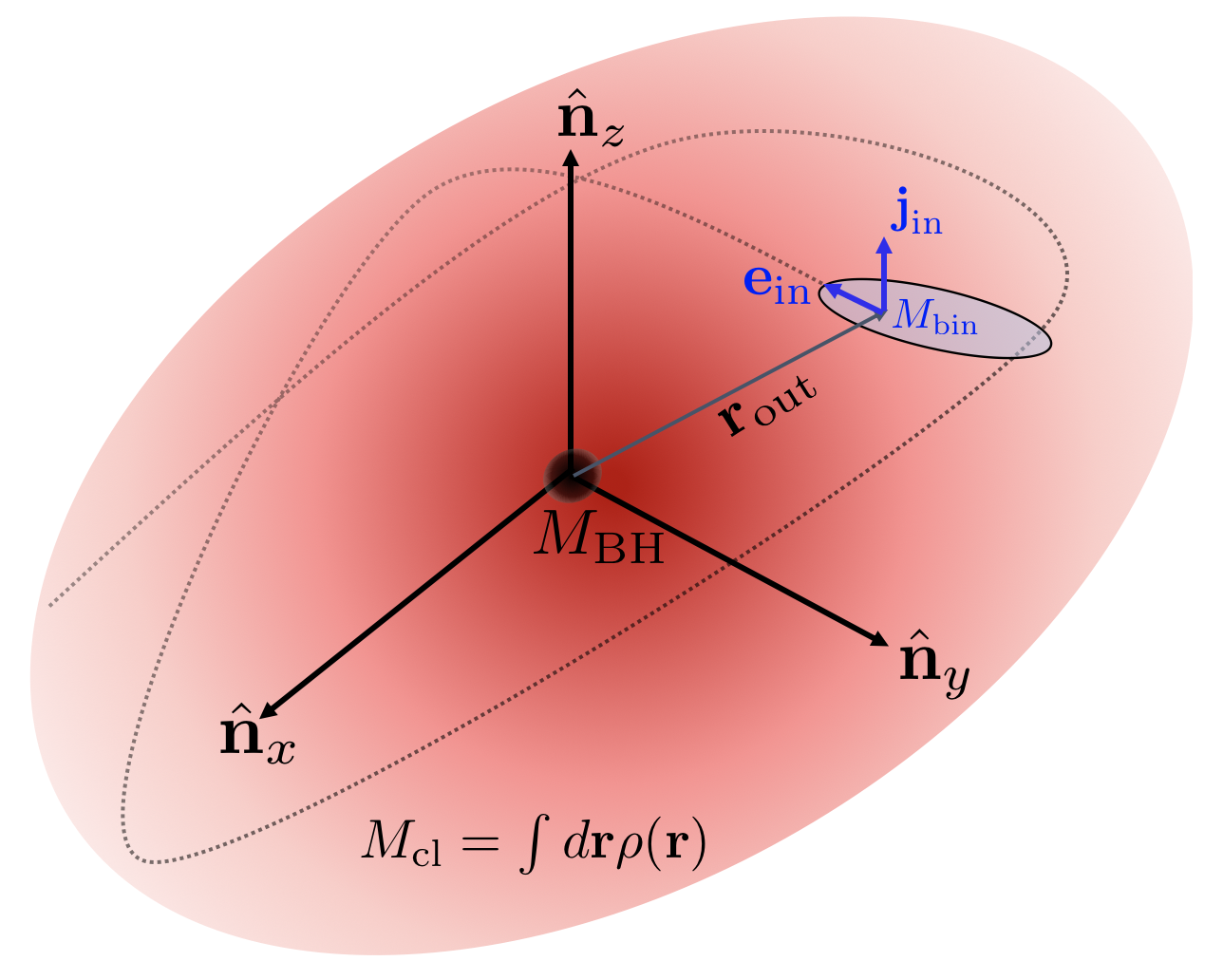}
    \caption{Coordinate system. The fixed Cartesian system with origin at the SMBH defines the symmetry axes of the ellipsoidal cluster density profile $\rho(\rvec)$. The position of the binary's barycenter is $\rvec_{\rm out}$, and relative to this location we define the eccentricity and specific angular momentum vectors, $\evec_{\rm in}$ and $\jvec_{\rm in}$,  to fully describe the binary's Keplerian orbit with semi-major axis $a_{\rm in}$.}
    \label{fig:coordinates}
\end{figure}

We consider a stellar binary system of total mass $M_\mathrm{bin}$ with semi-major axis $a_\mathrm{in}$, orbiting a central SMBH of mass $M_\mathrm{BH}$ with semi-major axis $a_\mathrm{out}$, and embedded within a nuclear star cluster with mass density $\rho(\rvec)$. To model the orbit of the inner binary system, we follow the vectorial formalism \citep[e.g.,][]{tremaine2009, tremaine2014}. Here, the system is characterized by the vectors
\begin{equation}
\label{eq:vectors}
    \evec \equiv e \, \ehat \hspace{2em} \jvec \equiv \sqrt{1 - e^2} \, \jhat
\end{equation}
where $\ehat$ points toward the pericenter of the inner binary, $\jhat$ is parallel to the angular momentum vector, and $e$ is the eccentricity. We also introduce a third unit vector, $\qhat = \jhat \times \ehat$ to complete the coordinate system. Finally, we define Cartesian unit vectors $\nhat_x$, $\nhat_y$, and $\nhat_z$, which provide the reference frame with respect to the SMBH. A schematic diagram of the coordinate system is shown in Figure \ref{fig:coordinates}, and a full summary of the notation used in this paper is given in Table \ref{tab:notation}.

\begin{table}[ht]
\begin{tabularx}{\linewidth}{lX}
    \hline
    Symbol                          & Description                                                   \\ \hline
    $M_\mathrm{BH}$                 & Mass of the central SMBH                                      \\
    $M_\mathrm{bin}$                & Total mass of the inner binary                                \\
    $M_\mathrm{cl}$                 & Total mass of the nuclear star cluster                        \\
    $\rvec_\mathrm{in}$             & Displacement vector of the inner binary                       \\
    $\rvec_\mathrm{out}$, $\rvec$   & Position vector of the outer binary                           \\
    $a_\mathrm{in}$, $a$            & Semi-major axis of the inner binary                           \\
    $a_\mathrm{out}$                & Semi-major axis of the outer binary                           \\
    $\jvec_\mathrm{in}$, $\jvec$    & Normalized angular momentum vector of the inner binary        \\
    $\jvec_\mathrm{out}$            & Normalized angular momentum vector of the outer binary        \\
    $\evec_\mathrm{in}$, $\evec$    & Eccentricity vector of the inner binary                       \\
    $e_\mathrm{in}$, $e$            & Eccentricity of the inner binary                              \\
    $e_\mathrm{out}$                & Eccentricity of the outer binary                              \\
    $i_\mathrm{in}$, $i$            & Inclination of the inner binary relative to the $z$-axis      \\
    $i_\mathrm{out}$                & Inclination of the outer binary relative to the $z$-axis      \\
    $\omega_\mathrm{in}$, $\omega$  & Argument of pericenter of the inner binary                    \\
    $\Omega_\mathrm{in}$, $\Omega$  & Longitude of the ascending node of the inner binary           \\
    $\Omega_\mathrm{out}$           & Longitude of the ascending node of the outer binary           \\
    $\nhat_i$                       & $i^\mathrm{th}$ Cartesian unit vector                         \\
    $\Phi$                          & Gravitational potential                                       \\ 
    $\Phi_\mathrm{cl}$              & Cluster potential                                             \\
    $\Phi_{ij}$                     & Tidal tensor, $\partial^2 \Phi / \partial x_i \partial x_j$   \\
    $\rho$                          & Mass density of the nuclear star cluster                      \\
    \hline
\end{tabularx}
\caption{Summary of the notation used throughout this paper.}
\label{tab:notation}
\end{table}

\subsection{Equations of Motion} \label{sec:equations_of_motion}
Given a smooth potential $\Phi$ that changes over scales much greater than that of the inner binary separation (i.e., $|d \log(\Phi) / dr|^{-1} \ll a_\mathrm{in}$), we may Taylor-expand the potential about some position $\rvec$, assuming that only the tidal field is important:
\begin{align}
\begin{split}
\label{eq:taylor}
    \Phi &\approx \frac{x^2}{2} \frac{\partial^2 \Phi}{\partial x^2} \bigg|_{\rvec} + \frac{y^2}{2} \frac{\partial^2 \Phi}{\partial y^2} \bigg|_{\rvec} + \frac{z^2}{2} \frac{\partial^2 \Phi}{\partial z^2} \bigg|_{\rvec}  \\
    &\quad + xy \frac{\partial^2 \Phi}{\partial x\partial y}\bigg|_{\rvec}
    + xz \frac{\partial^2 \Phi}{\partial x\partial z}\bigg|_{\rvec}
    + yz \frac{\partial^2 \Phi}{\partial y\partial z}\bigg|_{\rvec}\\
    &= \frac{1}{2} \sum_{i,j=x,y,z} \Phi_{ij}(\rvec) \, x_i x_j.
\end{split}
\end{align}
In this particular application, $\rvec = \rvec_\mathrm{out}$ gives the position of the barycenter of the inner binary relative to the SMBH, and $x_i = \nhat_i \cdot \rvec_\mathrm{in}$ gives the components of the displacement vector of the inner binary.

If we further assume that both the tidal potential $\Phi_{ij}$ and the inner angular momentum change at timescales much longer than the period of the binary, then we may average the tidal potential over the orbit of the inner binary. As demonstrated in Appendix \ref{ap:equations_of_motion}, the time-averaged potential reads
\begin{equation}
\label{eq:averaged_pot}
\begin{multlined}
    \langle \Phi \rangle = \frac{a_\mathrm{in}^2}{4} \sum_{i,j=x,y,z} \Phi_{ij}(\rvec) \big[ 5(\nhat_i \cdot \evec)(\nhat_j \cdot \evec) \\ - (\nhat_i \cdot \jvec)(\nhat_j \cdot \jvec) + j^2 \, \delta_{ij} \big].
\end{multlined}
\end{equation}
The equations of motion for the inner binary system are then given by
\begin{gather}
    \begin{multlined}
        \frac{d \jvec}{dt} = \frac{a_\mathrm{in}^{3/2}}{2 \sqrt{G M_\mathrm{bin}}} \sum_{i,j=x,y,z} \Phi_{ij}(\rvec) \big[ (\nhat_j \cdot \jvec)(\jvec \times \nhat_i) \\ - 5(\nhat_j \cdot \evec)(\evec \times \nhat_i) \big]
    \end{multlined} \label{eq:djdt} \\
    \begin{multlined} 
        \frac{d \evec}{dt} = \frac{a_\mathrm{in}^{3/2}}{2 \sqrt{G M_\mathrm{bin}}} \sum_{i,j=x,y,z} \Phi_{ij}(\rvec) \big[ (\nhat_j \cdot \jvec)(\evec \times \nhat_i) \\ - 5(\nhat_j \cdot \evec)(\jvec \times \nhat_i)  + \delta_{ij} \, (\jvec \times \evec) \big].
    \end{multlined} \label{eq:dedt}
\end{gather}
The full derivations for these equations can be found in Appendix \ref{ap:equations_of_motion}.

In addition to Equations (\ref{eq:djdt}) and (\ref{eq:dedt}), we also include the relativistic precession of the $\evec$ vector, which adds an additional term given by
\begin{equation}
\label{eq:gr_precession}
    \frac{d \evec}{dt} = \frac{\dot{\omega}_{\rm GR}}{(1 - e^2)^{3/2}}\, \jvec \times \evec
\end{equation}
where
\begin{equation}
\label{eq:omega_gr}
\dot{\omega}_{\rm GR}=\frac{3 G^{3/2} M_\mathrm{bin}^{3/2}}{a_\mathrm{in}^{5/2} c^2 }. 
\end{equation}

\subsection{Cluster Model} \label{sec:cluster}
Although our treatment is general for a wide range of density profiles, we shall focus our attention to one specific family of density profiles known as the triaxial $\gamma$-family, which is given by
\begin{equation}
\label{eq:gamma_family}
    \rho(x, y, z) = \frac{(3 - \gamma) M_\mathrm{cl}}{4 \pi b c} \frac{s}{m^\gamma (m + s)^{4 - \gamma}}
\end{equation}
where $M_\mathrm{cl}$ is the total mass of the cluster, $s$ is the scale radius, and $m^2 = x^2 + y^2/b^2 + z^2/c^2$. For reference, the mass enclosed by the density profile in the region $m < \widetilde{m}$ is
\begin{align}
\begin{split}
\label{eq:mass_enclosed}
    M_\mathrm{encl}(m < \widetilde{m}) &= 4 \pi b c \int_0^{\widetilde{m}} m^2 \rho(m) \, dm \\
    &= M_\mathrm{cl} \left( \frac{\widetilde{m}}{\widetilde{m} + s} \right)^{3 - \gamma}.
\end{split}
\end{align}

Motivated by the observational results of \cite{chatzo2015}, throughout the remainder of this paper we will assume that $s = 4 \, \mathrm{pc}$, and normalize the cluster potential such that the mass enclosed by the cluster in the scale radius is given by $M_\mathrm{encl}(m < 4 \, \mathrm{pc}) = 2 \, M_\mathrm{BH}$. Thus, $M_\mathrm{cl} = 2^{4 - \gamma} \, M_\mathrm{BH}$.

\subsubsection{Potential}

The potential associated with this mass distribution has no explicit form and has to be calculated by numerically solving a one-dimensional integral. However, there are limiting cases that allow for explicit and simple expressions that will prove useful for our theoretical analysis in Sections \ref{sec:torus-filling} and \ref{sec:non-torus-filling}.

First, in the limiting case of a spherically symmetric potential ($b = c = 1$, or $m=r$), the associated potential for the $\gamma$-family can be written analytically as
\begin{equation}
\label{eq:gamma_pot_spherical}
    \Phi_\mathrm{cl}(r) = -\frac{GM_\mathrm{cl}}{s (2 - \gamma)} \left[ 1 - \left( \frac{r}{r + s} \right)^{2 - \gamma} \right]
\end{equation}
in the case $\gamma \neq 2$ \citep{renaud2010}. This, in turn, gives an expression for the tidal tensor as
\begin{equation}
\label{eq:gamma_tt_spherical}
    \Phi_{\mathrm{cl}}^{ij} = \frac{GM_\mathrm{cl}}{r^\gamma (r + s)^{3 - \gamma}} \left[ \delta_{ij} - x_i x_j \frac{3r + s \gamma}{(r + s)r^2} \right].
\end{equation}

Second, in the limiting case of a slightly flattened density distribution, $b=1$ and $1-c^2\ll 1$, we show in Appendix \ref{ap:flattened_hernquist} that an explicit form of the potential can be found using a quadrupolar expansion. In particular, for a Hernquist potential ($\gamma=1$), we expand the density distribution in spherical coordinates as
\begin{equation}
\rho(r,\theta)= \frac{M_\mathrm{cl}}{2 \pi c} \frac{s}{r (r + s)^3}
\left[1 -\epsilon_z \left(\frac{4r+s}{r+s}\right)\cos^2\theta
\right]
\end{equation}
with $\epsilon_z=(1-c^2)/c^2<1/2$. The full potential is given by Equation (\ref{eq:hernquist_flat}) and an approximated solution for $r<s$ is
\begin{equation}
\label{eq:hernquist_flat_approx}
    \Phi_\mathrm{cl}(r,\theta)\approx -\frac{GM_\mathrm{cl}}{c(r+s)}\left[1-\frac{\epsilon_z}{3}+
    \frac{13\epsilon_z}{24}\frac{r}{s} -\frac{\epsilon_z }{6}
    \frac{r}{s}\cos^2 \theta\right].
 \end{equation}  

\subsubsection{Velocity Dispersion and Distribution Function}
For simplicity, we assume spherical symmetry when estimating the velocity dispersion and distribution function for our potential model. Thus, the velocity dispersion can be obtained from the Jeans equations as
\begin{equation}
    \sigma^2(r) = \frac{1}{\rho(r)} \int_r^\infty \frac{M_\mathrm{encl}(r') \rho(r')}{r'^2} \, dr'
\end{equation}
\citep{binney+tremaine}. Similarly, the distribution function can be obtained from Eddington's formula as
\begin{equation}
    f(\mathcal{E}) = \frac{1}{\sqrt{8}\pi^2} \left[ \int_0^\mathcal{E} \frac{d \Psi}{\sqrt{\mathcal{E} - \Psi}} \frac{d^2 \nu}{d\Psi^2} + \frac{1}{\sqrt{\mathcal{E}}} \frac{d \nu}{d \Psi} \bigg|_{\Psi = 0} \right]
\end{equation}
where $\Psi \equiv -\Phi$ is the relative potential, $\mathcal{E} \equiv \Psi - \tfrac{1}{2} v^2$ is the relative energy, and $\nu(r) \equiv \rho(r) / M_\mathrm{cl}$ is the spatial probability density \citep{binney+tremaine}. Specific details regarding our implementation of these formulae as well as an analytic expression for the velocity dispersion of a Hernquist-profile cluster can be found in Appendix \ref{ap:df}.

\subsection{Simulation Procedure} \label{sec:simulation}
In order to evolve the orbit of the binary system numerically, we have developed a hybrid Python code to evolve both the barycenter position $\rvec_\mathrm{out}$ of the system, and to compute the singly-averaged tidal torque on the inner binary in an arbitrary potential. The code, which is available on github\footnote{\url{https://github.com/mwbub/binary-evolution}}, utilizes \texttt{galpy} \citep{bovy2015} to take advantage of its large library of potentials. With this code, the simulation procedure is as follows:
\begin{enumerate}
    \item Integrate the orbit of the barycenter, $\rvec_\mathrm{out}$,
    about the Galactic center using \texttt{galpy}.
    \item At each time step, compute the tidal tensor $\Phi_{ij}$ of the combined black hole plus cluster potential.
    \item Evolve the $\jvec$ and $\evec$ vectors according to Equations (\ref{eq:djdt}) and (\ref{eq:dedt}), respectively.
\end{enumerate}

The computation of the tidal tensor is also accomplished via \texttt{galpy}. This allows the code to be applied to a wide variety of potentials, including many triaxial potentials as well as arbitrary sums of potentials. In particular, we use \texttt{galpy}'s \texttt{KeplerPotential} and \texttt{TwoPowerTriaxialPotential} to compute the tidal tensor for the sum of the SMBH Keplerian potential and the $\gamma$-family cluster potential, respectively.

In all of our simulations, we integrate for approximately 1,000 secular timescales, up to a maximum of $1 \, \mathrm{Gyr}$. For simplicity, we compute the secular timescale here with Equation (\ref{eq:secular_timescale_spherical}) rather than Equation (\ref{eq:secular_timescale}).

\section{Torus-Filling Dynamics in Axisymmetric Potentials} \label{sec:torus-filling}
In this section, we examine the dynamics of binaries whose outer orbits densely fill an axisymmetric torus on timescales shorter than those at which the inner binary is torqued by the tidal field. This problem was studied in detail by \cite{HR2019a, HR2019b}, who develop a general formalism to describe the dynamics of torus-filling binaries in axisymmetric potentials, generalizing previous results regarding the effect of galactic tides on stellar binaries \citep[e.g.,][]{HT1986}. Here, we demonstrate that our simulations are consistent with the results of \cite{HR2019a, HR2019b}, and examine the additional effect of an SMBH on these considerations.

\subsection{Torus-Averaged Equations} \label{sec:torus_equations}
When dealing with torus-filling orbits in axisymmetric potentials, one can approximate the the evolution of the inner binary by averaging $\Phi_{ij}$ over many periods of the outer binary, a technique which we refer to as ``torus-averaging". Note that this term is distinguished from the commonly-employed term ``double-averaging" for the purposes of this work. Here, we define double-averaging as averaging over the dynamical timescale of the outer binary. By contrast, torus-averaging refers to averaging over many dynamical timescales such that the outer orbit may densely fill an axisymmetric torus. For this work, this is in practice equivalent to averaging over the timescale of the nodal precession of the outer orbit. The regimes in which these two forms of averaging break down are distinct, as shown in Figure \ref{fig:regimes} and discussed in Section \ref{sec:non-torus-filling}.

As shown by \cite{HR2019a}, in this torus-averaged limit the tidal tensor has only diagonal terms, and $\langle \Phi_{xx} \rangle = \langle \Phi_{yy} \rangle$. In this case, the torus-averaged potential reads
\begin{align}
\begin{split}
\label{eq:doubly_averaged_pot}
    \langle \langle \Phi \rangle \rangle &= \frac{a_\mathrm{in}^2}{4} \sum_{i=x,y,z} \langle \Phi_{ii} \rangle \left[ 5 (\nhat_i \cdot \evec)^2 - (\nhat_i \cdot \jvec)^2 + j^2 \right] \\
    &= \frac{3a_\mathrm{in}^2}{2} \langle \Phi_{zz} + \Phi_{xx} \rangle \left[ \tfrac{1}{2} \Gamma (5e_z^2 - j_z^2) + \tfrac{1}{4} e^2 (1 - 5 \Gamma) \right]
\end{split}
\end{align}
where
\begin{equation}
\label{eq:Gamma}
    \Gamma \equiv \frac{ \langle \Phi_{zz} - \Phi_{xx} \rangle }{3 \langle \Phi_{zz} + \Phi_{xx} \rangle}.
\end{equation}
For an SMBH alone, the potential reduces to the Keplerian case where $\Gamma = 1$, recovering the well-known Lidov-Kozai potential.

In the torus-averaged potential, $j_z = \sqrt{1 - e^2} \cos i_\mathrm{in}$ is a constant of motion and the potential is integrable. Using the conservation of $j_z$ and the secular energy, \cite{HR2019b} show that a nearly circular inner orbit with $j_z \approx \cos i_0$ reaches a maximum eccentricity given by
\begin{equation}
\label{eq:e_max}
    e_\mathrm{max} = \sqrt{1 - \frac{10 \Gamma}{1 + 5 \Gamma} \cos^2 i_0}.
\end{equation}
This result is valid for orbits with $\Gamma > 1/5$, where the phase space structure resembles that of the Lidov-Kozai potential in which $\omega$ undergoes libration. For $0 < \Gamma < 1/5$, no libration of the argument of pericenter is found. In the case $\Gamma = 1$, the formula reduces to the previously known $e_\mathrm{max} = \sqrt{1 - \tfrac{5}{3} \cos^2 i_0}$ \citep[e.g.,][]{lidov1962, kozai1962}. As we will see in Section \ref{sec:sim_comparison}, the presence of an SMBH conspires to keep $\Gamma > 1/5$ in most regimes. We also stress that Equation (\ref{eq:e_max}) is valid only for inner orbits which are initially nearly circular, and as such we only use this expression for demonstrative purposes.

From the equations of motion for $\jvec$ and $\evec$, the characteristic secular timescale for the evolution in the torus-averaged potential is given by
\begin{equation}
\label{eq:secular_timescale}
    \tau_\mathrm{sec}^{-1} = \frac{3 a_\mathrm{in}^{3/2}}{2 \sqrt{GM_\mathrm{bin}}} \langle \Phi_{zz} + \Phi_{xx} \rangle.
\end{equation}
Note that this expression is equivalent to Equation (34) of \citet{HR2019b}.

In the special case of a circular orbit in the Galactic midplane, embedded within a spherical $\gamma$-family cluster with a central SMBH, the torus-averaged tidal tensor can be derived from Equation (\ref{eq:gamma_tt_spherical}) as
\begin{align}
\begin{split}
\label{eq:gamma_tt_averaged}
    \langle \Phi_{xx} \rangle &= \frac{G M_\mathrm{cl}}{a_\mathrm{out}^\gamma (a_\mathrm{out} + s)^{3 - \gamma}} \left[ 1 - \frac{3 a_\mathrm{out} + s \gamma}{2 (a_\mathrm{out} + s)} \right] - \frac{G M_\mathrm{BH}}{2 a_\mathrm{out}^3} \\ 
    \langle \Phi_{zz} \rangle &= \frac{G M_\mathrm{cl}}{a_\mathrm{out}^\gamma (a_\mathrm{out} + s)^{3 - \gamma}} + \frac{G M_\mathrm{BH}}{a_\mathrm{out}^3}.
\end{split}
\end{align}
In turn, $\Gamma$ becomes
\begin{equation}
\label{eq:Gamma_spherical}
    \Gamma = \frac{3M_\mathrm{BH} +  M_\mathrm{cl} \, a_\mathrm{out}^{3 - \gamma}(a_\mathrm{out} + s)^{\gamma - 4}(3a_\mathrm{out} + s \gamma)}{3[M_\mathrm{BH} + M_\mathrm{cl} \, a_\mathrm{out}^{3 - \gamma}(a_\mathrm{out} + s)^{\gamma - 4}(a_\mathrm{out} + s(4 - \gamma))]}.
\end{equation}
As expected, this expression reduces to $\Gamma = 1$ for $M_\mathrm{cl} = 0$. The expression also approaches unity asymptotically as $a_\mathrm{out} \to \infty$ and the cluster potential appears increasingly Keplerian. The secular timescale then becomes
\begin{equation}
\label{eq:secular_timescale_spherical}
\begin{multlined}
    \tau_\mathrm{sec}^{-1} = \frac{3 a_\mathrm{in}^{3/2}}{2 \sqrt{GM_\mathrm{bin}}} \bigg( \frac{G M_\mathrm{BH}}{2 a_\mathrm{out}^3} \\ + \frac{G M_\mathrm{cl}}{a_\mathrm{out}^\gamma (a_\mathrm{out} + s)^{3 - \gamma}} \left[ 2 - \frac{3 a_\mathrm{out} + s \gamma}{2 (a_\mathrm{out} + s)} \right] \bigg).
\end{multlined}
\end{equation}
For elliptical orbits, one can use this expression to approximate the secular timescale by modifying $a_\mathrm{out} \to a_\mathrm{out} \sqrt{1 - e_\mathrm{out}^2}$ \citep[e.g.,][]{petrovich2017}.

\subsection{Comparison with Simulations} \label{sec:sim_comparison}

\begin{figure}[ht]
    \centering
    \includegraphics[width=\linewidth]{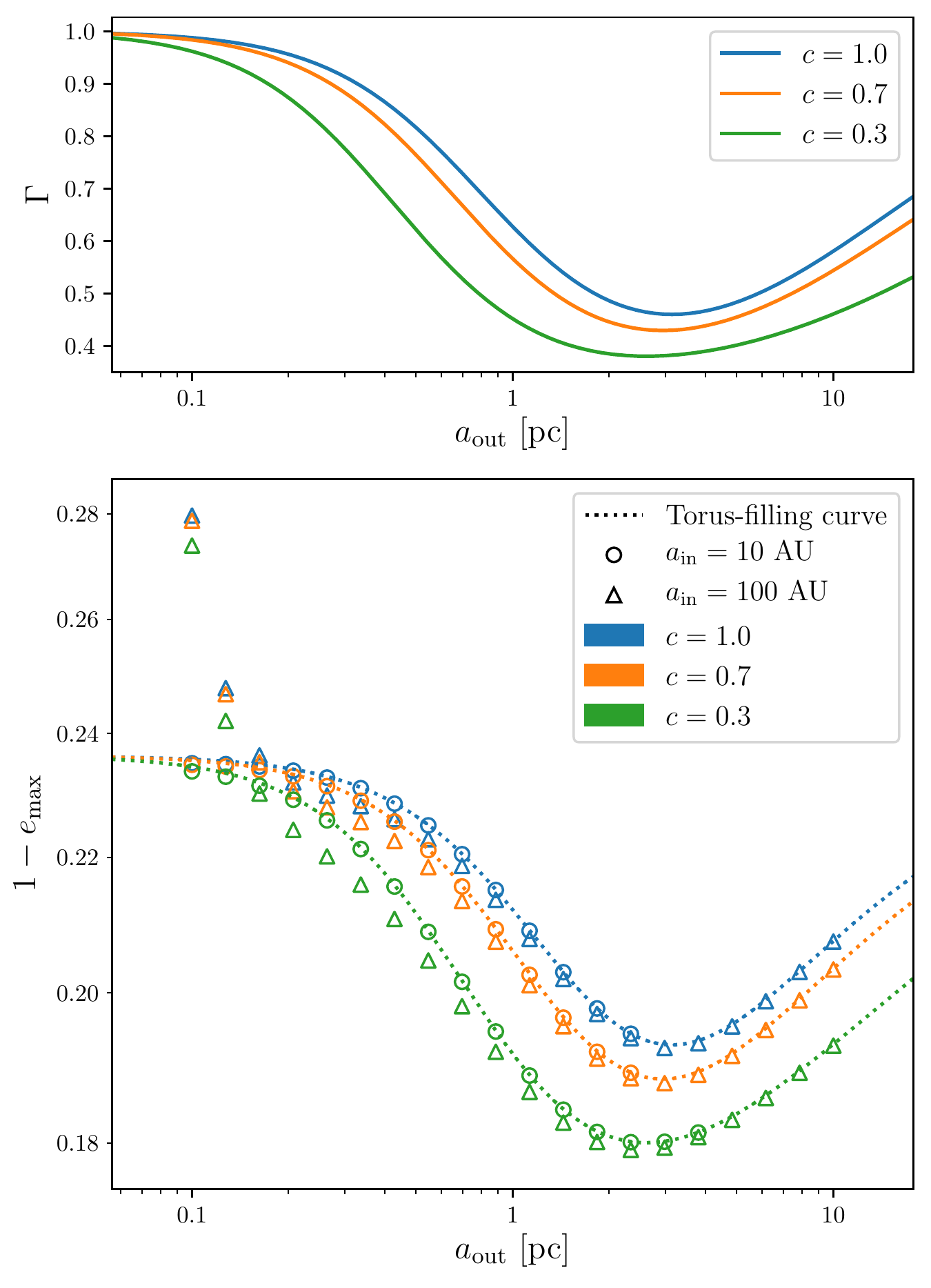}
    \caption{$\Gamma$ and $1 - e_\mathrm{max}$ as a function of $a_\mathrm{out}$ for circular orbits in the Galactic midplane, embedded within an axisymmetric Hernquist-profile ($\gamma = 1$) cluster potential with various flattening factors $c$. \textbf{Top panel:} $\Gamma$ as a function of $a_\mathrm{out}$. The $c=1.0$ curve is generated analytically via Equation (\ref{eq:Gamma_spherical}), whereas the $c=0.7$ and $c=0.3$ curves are generated numerically via Equation (\ref{eq:Gamma}). \textbf{Bottom panel:} $1 - e_\mathrm{max}$ as a function of $a_\mathrm{out}$, assuming an initial inclination of $i_0 = 60^\circ$ for the inner binary. The dashed lines represent the analytic prediction given by Equation (\ref{eq:e_max}). The circle and triangle markers represent the numerical results from the singly-averaged code, run with $a_\mathrm{in} = 10 \, \mathrm{AU}$ and $100 \, \mathrm{AU}$.}
    \label{fig:gamma_emax}
\end{figure}

The predictions of the torus-averaged equations hold in our singly-averaged code, so long as the secular timescale is much longer than the timescale to fill an axisymmetric torus. This can be seen in Figure \ref{fig:gamma_emax}, which compares the analytic $e_\mathrm{max}$ given by Equation (\ref{eq:e_max}) with the numerical results from our singly-averaged code. Here, we perform simulations of the secular evolution for circular orbits in the Galactic midplane, which trivially fill a torus, using an inner semi-major axis of both $a_\mathrm{in} = 10 \, \mathrm{AU}$ and $100 \, \mathrm{AU}$. The figure also demonstrates the effect of varying the flattening parameter $c$ of the potential, defined in Equation (\ref{eq:gamma_family}). Note that for demonstration purposes, these particular simulations omit the quenching due to relativistic precession. In addition, there are no data points for $a_\mathrm{out} > 3 \, \mathrm{pc}$ in the $a_\mathrm{in} = 10 \, \mathrm{AU}$ case, as beyond this point the secular timescale exceeds our maximum integration time of $1 \, \mathrm{Gyr}$.

The numerical $e_\mathrm{max}$ results in the $a_\mathrm{in} = 10 \, \mathrm{AU}$ case are in very good agreement with the analytic predictions at all values of $a_\mathrm{out}$. In the $a_\mathrm{in} = 100 \, \mathrm{AU}$ case, the agreement is also good for $a_\mathrm{out} \gtrsim 1 \, \mathrm{pc}$, but begins to diverge from the analytic curve at smaller distances from the SMBH. This divergence occurs as the dynamical timescale of the outer binary becomes comparable to the secular timescale. In Figure \ref{fig:regimes}, this corresponds to the orange-shaded region where double-averaging breaks down.

The effect of the SMBH is also clear from this figure. At small values of $a_\mathrm{out}$, the central black hole dominates and the system approximates isolated three-body dynamics. This corresponds to $\Gamma = 1$. At larger $a_\mathrm{out}$, the influence of the cluster is more apparent, causing a dip in the value of $\Gamma$. However, the presence of the SMBH keeps $\Gamma$ greater than the critical value of $1/5$ at all distances, in contrast to the behavior seen in \cite{HR2019a}. At large values of $a_\mathrm{out}$, $\Gamma$ begins to converge again toward unity, as expected. 

\begin{figure*}[ht]
    \centering
    \includegraphics[width=\linewidth]{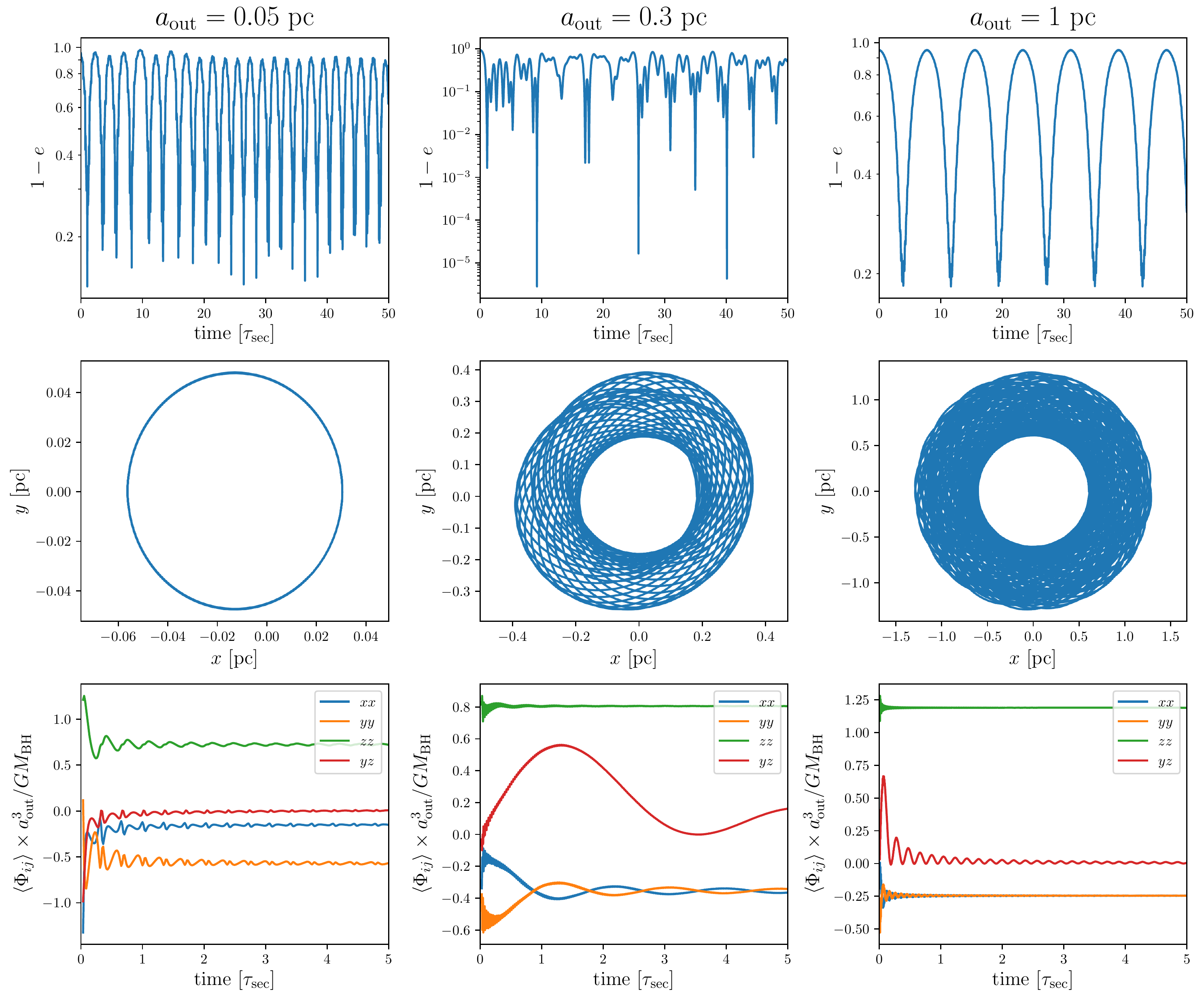}
    \caption{Evolution of binaries in three different dynamical regimes. Each binary is embedded within an axisymmetric Hernquist-profile cluster potential taken with $c = 0.7$. The outer orbits are each given inclination $i_\mathrm{out} = 30^\circ$ and eccentricity $e_\mathrm{out} = 0.3$. The inner orbits are each given semi-major axis $a_\mathrm{in} = 30 \, \mathrm{AU}$ and initial inclination $i_0 = 60^\circ$ relative to the $z$-axis. \textbf{Left column:} Quasi-Keplerian orbit (Section \ref{sec:keplerian}). \textbf{Middle column:} Chaotic non-torus-filling orbit (Section \ref{sec:chaos}). \textbf{Right column:} Torus-filling orbit (Section \ref{sec:torus-filling}). \textbf{Top row:} $1 - e$ as a function of time over 50 secular timescales. \textbf{Middle row:} $x$-$y$ projection of the outer orbit over one secular timescale. \textbf{Bottom row:} Averaged tidal tensor components as a function of time over five secular timescales.}
    \label{fig:e_xy_tt}
\end{figure*}

The right column of Figure \ref{fig:e_xy_tt} gives an example of the evolution of a torus-filling binary. Here, the center-right panel shows that the binary densely fills a torus in a single secular timescale. In addition, the lower-right panel demonstrates that the $\langle \Phi_{xx} \rangle$ and $\langle \Phi_{yy} \rangle$ components of the averaged tidal tensor converge to each other, and that the cross terms of the tidal tensor vanish within a secular timescale. As such, we observe regular eccentricity cycles with a well-defined $e_\mathrm{max}$ in the upper-right panel, which is in good agreement with the torus-averaged predictions.

\section{Non-Torus-Filling Dynamics} \label{sec:non-torus-filling}

In this section, we discuss the dynamics of binaries in axisymmetric potentials whose outer orbits fail to densely fill a torus within the secular timescale. In this regime, the analytic formalism of \cite{HR2019a, HR2019b} breaks down, and we observe substantially different behavior. Most interestingly, this regime can give rise to secular chaos in the evolution of the inner orbit, as previously studied by \cite{petrovich2017}. This can in turn induce extreme eccentricities in the inner binary, and consequently greatly enhanced merger rates.

\subsection{Nodal Precession Timescale} \label{sec:timescale}
The location of the non-torus-filling regime can be estimated by considering the timescale of the nodal precession of the outer orbit compared to the secular timescale. Therefore, here we calculate the nodal precession rate due to an axisymmetric Hernquist potential and SMBH. We ignore the effect of apsidal precession in this analysis because it has a less significant dynamical role: eccentric orbits that fail to fill a torus due to slow apsidal precession alone still produce an effectively axisymmetric potential, as the binary evolution is dominated by the SMBH (i.e., the quadrupolar Lidov-Kozai potential).

We average the approximate cluster potential from Equation (\ref{eq:hernquist_flat_approx}) over one dynamical timescale of the outer binary. Thus, for a circular orbit we get
\begin{equation}
\label{eq:hernquist_averaged}
    \langle \Phi_\mathrm{cl} \rangle \approx \frac{\epsilon_z }{6c}\frac{GM_\mathrm{cl}}{s}
    \frac{a_\mathrm{out}}{s}(\jvec_\mathrm{out}\cdot \nhat_z)^2+\mbox{cst.}
  \end{equation}  
and the orientation of $\jvec_\mathrm{out}$ is given by
\begin{equation}
\label{eq:dj_out}
    \frac{d \jvec_\mathrm{out}}{dt} = \dot{\Omega}_\mathrm{out}   (\nhat_z \cdot \jvec_\mathrm{out})(\jvec_\mathrm{out} \times \nhat_z)
\end{equation}
where
\begin{equation}
\label{eq:Omega_dot}
    \dot{\Omega}_\mathrm{out}=\frac{\epsilon_z}{3c}\left(\frac{GM_\mathrm{cl}}{s^3}\right)^{1/2} \left(\frac{a_\mathrm{out}}{s}\right)^{1/2}
    \left( \frac{M_\mathrm{cl}}{M_{\rm BH}}\right)^{1/2}.
\end{equation}
This precession rate has to be compared to the secular timescale in Equation (\ref{eq:secular_timescale}) to determine the dynamical regime of the system. For orbits inside $\sim 1 \, \mathrm{pc}$, the secular timescale is dominated by the black hole and we can write
\begin{equation}
\label{eq:tau_Omega_dot}
    \tau_{\rm sec}\dot{\Omega}_\mathrm{out}=
    \frac{4\epsilon_z}{9c}\left(\frac{M_\mathrm{bin}^{1/2}
    M_\mathrm{cl}}{M_\mathrm{BH}^{3/2}}\right)
    \left(\frac{a_\mathrm{out}^{7/2}}
    {a_\mathrm{in}^{3/2}s^2}\right).
\end{equation}
For our fiducial parameters $s=4 \, \mathrm{pc}$, $M_\mathrm{cl}=8 \, M_{\rm BH}$, and $M_\mathrm{bin}=10 \, M_\odot$ we get
\begin{equation}
\label{eq:tau_Omega_dot_subbed}
    \tau_{\rm sec}\dot{\Omega}_\mathrm{out} \approx 0.7
    \left(\frac{\epsilon_z/c}{0.2}\right)
    \left(\frac{a_\mathrm{out}}{0.2 \, \mbox{pc}}\right)^{7/2}
    \left(\frac{10 \, \mbox{AU}}{a_\mathrm{in}}\right)^{3/2}.
\end{equation}
Recall that this expression is derived for small $\epsilon_z$ and is only physically valid for $\epsilon_z=(1/c^2-1)<1/2$, or $c>\sqrt{2/3} \approx 0.81$ (see Appendix \ref{ap:flattened_hernquist}). Nevertheless, we shall still use this expression for smaller $c$ below to provide guidance on the typical dynamical regime.

The relevant dynamical regimes can now be more precisely defined as follows: for $\tau_\mathrm{sec} \dot{\Omega}_\mathrm{out} \gg 1$, orbits are torus-filling and the considerations of Section \ref{sec:torus-filling} are valid; for $\tau_\mathrm{sec} \dot{\Omega}_\mathrm{out} \ll 1$, orbits are nearly Keplerian, as discussed in Section \ref{sec:keplerian} below; and for $\tau_\mathrm{sec} \dot{\Omega}_\mathrm{out} \approx 1$, secular chaos is induced, as discussed in Section \ref{sec:chaos}.

\subsection{Quasi-Keplerian Orbits} \label{sec:keplerian}
In the limit $\tau_\mathrm{sec} \dot{\Omega}_\mathrm{out} \ll 1$, orbits are approximately Keplerian on the secular timescale, and the isolated three-body dynamics of the SMBH and binary become valid. This can be seen in the left column of Figure \ref{fig:e_xy_tt}, which shows the evolution of a binary with $\tau_\mathrm{sec} \dot{\Omega}_\mathrm{out} \approx 10^{-2}$. The center-left panel shows that the orbit traces out a nearly closed ellipse in a secular timescale. In turn, we recover the standard Lidov-Kozai cycles, as shown in the upper-left panel. Note that the small secondary oscillations in each Lidov-Kozai cycle here are due to the breakdown of the double-averaging approximation as the secular timescale approaches the dynamical timescale.

In addition, the lower-left panel of Figure \ref{fig:e_xy_tt} shows that in the quasi-Keplerian case, $\langle \Phi_{xx} \rangle$ and $\langle \Phi_{yy} \rangle$ do not necessarily converge to each other as they do in the torus-filling case, for instance in the lower-right panel. Indeed, as shown by \cite{petrovich2017}, in this limit the symmetry axis of the nuclear star cluster is no longer the relevant frame for the dynamics of the inner binary, but rather it is the rotated frame of the outer orbital plane. Thus, the relative component $\jvec_\mathrm{in} \cdot \jhat_\mathrm{out}$ is conserved rather than $j_z$ as in the torus-filling case. Consequently, the maximum eccentricity is determined by the relative inclination of the inner binary to its outer orbital plane, rather than to the Galactic midplane.

\begin{figure}[t]
    \centering
    \includegraphics[width=\linewidth]{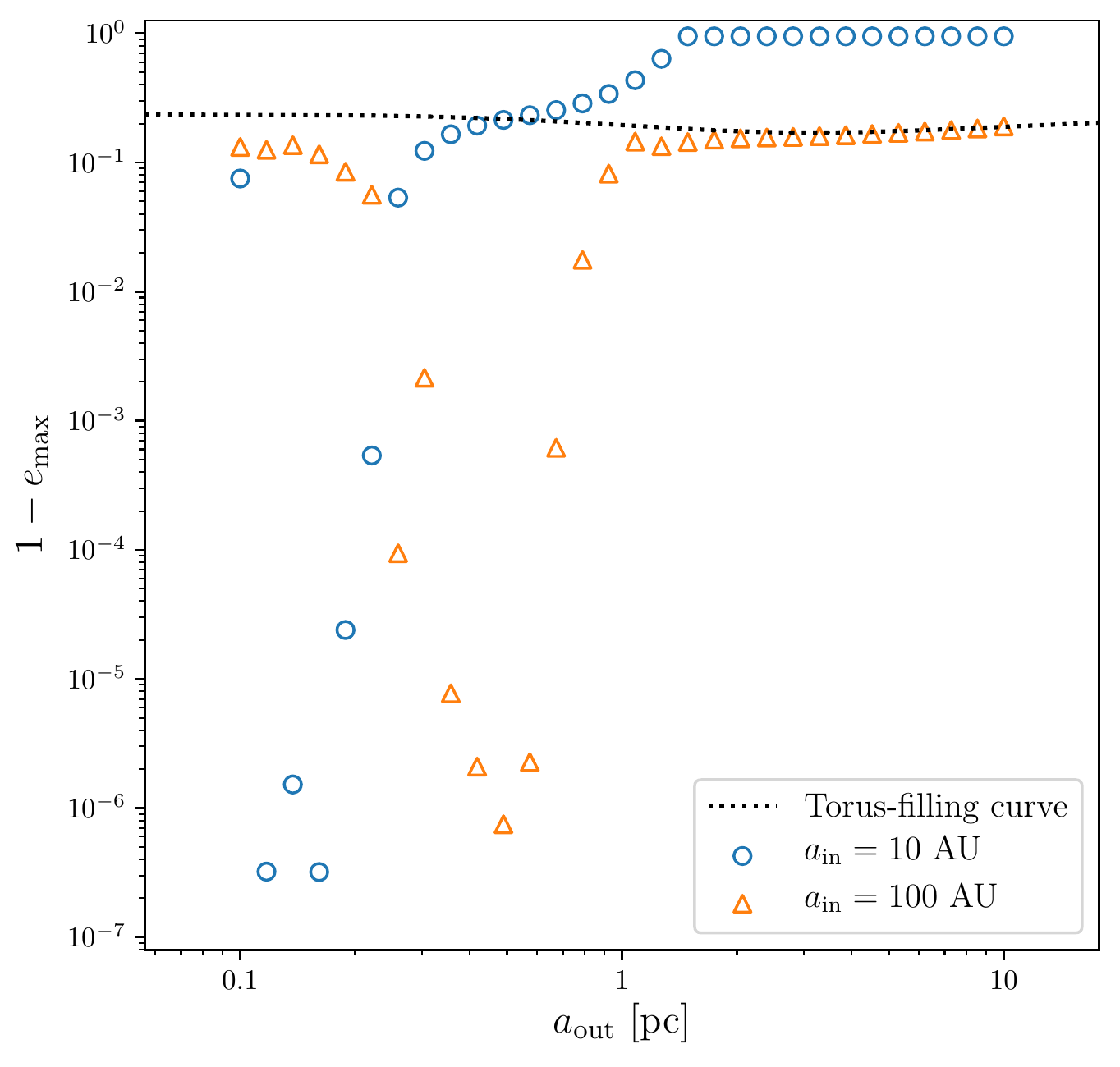}
    \caption{$1 - e_\mathrm{max}$ as a function of $a_\mathrm{out}$ for binaries in an axisymmetric Hernquist cluster potential taken with $c = 0.7$. The outer orbits are each taken to be circular and are given an inclination of $i_\mathrm{out} = 30^\circ$. The inner orbits are each given initial inclination $i_0 = 60^\circ$ relative to the $z$-axis, and are run with both $a_\mathrm{in} = 10 \, \mathrm{AU}$ and $100 \, \mathrm{AU}$. The quasi-Keplerian, chaotic, and torus-filling regimes are visible for both values of $a_\mathrm{in}$. In addition, the GR-quenching regime is visible for $a_\mathrm{in} = 10 \, \mathrm{AU}$ at larger values of $a_\mathrm{out}$.}
    \label{fig:emax_i30}
\end{figure}

\subsection{Secular Chaos} \label{sec:chaos}
In regions where the secular timescale is comparable to the nodal precession timescale, orbits are neither torus-filling nor Keplerian. In this case, the dynamical system becomes chaotic, and the inner orbit performs a random walk through the available parameter space \citep{petrovich2017}. As a consequence, the eccentricity of the inner binary can approach unity within a few secular timescales.

An example of this evolution is given in the middle column of Figure \ref{fig:e_xy_tt}, which demonstrates a binary with $\tau_\mathrm{sec} \dot{\Omega}_\mathrm{out} \approx 4$. Here, we observe from the center panel that the the orbit is certainly not Keplerian, but also fails to densely fill a torus within a secular timescale. Additionally, from the lower-center panel, we can see that the tidal tensor components do not converge to their torus-averaged values within a secular timescale. As a consequence, the evolution becomes chaotic, leading to the extreme eccentricities seen in the upper-center panel. Here, the eccentricity reaches levels of $1 - e < 10^{-5}$ within $10 \, \tau_\mathrm{sec}$.

A succinct summary of the various dynamical regimes is given by Figure \ref{fig:emax_i30}, which plots $1-e_\mathrm{max}$ as a function of $a_\mathrm{out}$ for binaries with inclined outer orbits and two different values of $a_\mathrm{in}$. At low values of $a_\mathrm{out}$, we can see that the binaries undergo modest eccentricity excitations consistent with the quasi-Keplerian regime. At slightly higher values of $a_\mathrm{out}$, the binaries begin to achieve extremely high eccentricities as they enter the chaotic regime. Toward the center of the chaotic regime, almost all binaries reach eccentricities of $1 - e < 10^{-4}$, a level which is relevant for mergers. Since $\tau_\mathrm{sec}$ is proportional to $a_\mathrm{in}^{-3/2}$, the chaotic regime is shifted to the left in the $a_\mathrm{in} = 10 \, \mathrm{AU}$ case as compared to the $100 \, \mathrm{AU}$ case. As $a_\mathrm{out}$ becomes larger still, the maximum eccentricities begin to converge to the torus-filling prediction, which is given by the black dotted line. At this point, relativistic precession begins to quench the eccentricity excitations for the $10 \, \mathrm{AU}$ binaries, until excitations are no longer observed for $a_\mathrm{out} \gtrsim 1 \, \mathrm{pc}$. In this same region, the $100 \, \mathrm{AU}$ binaries remain in good agreement with the torus-filling prediction.

Figure \ref{fig:emax_i30} is generated assuming that each outer orbit is circular. When the outer orbit is eccentric, as in Figure \ref{fig:e_xy_tt}, the dynamical regimes are qualitatively similar. The additional oscillations due to the apsidal precession of the outer orbit have the effect of extending the size of the chaotic regime in phase space, but otherwise do not lead to additional behavior. As discussed in Section \ref{sec:timescale}, apsidal precession alone is insufficient to excite extreme eccentricities in the inner binary. For very highly eccentric outer orbits, however, the breakdown of the double-averaging approximation as $\tau_\mathrm{sec}$ becomes small can lead to large eccentricities for highly-inclined inner orbits.

\section{Dynamics in Triaxial Potentials} \label{sec:triaxial}
In this section we discuss the influence of triaxiality on the evolution of Galactic-center binaries. As of yet, this regime has not been explored in the literature. Here, we provide an initial overview of these dynamics by describing the behavior in various regions of phase space, particularly as it contrasts with the axisymmetric case. 

We note that the dynamics in general triaxial potentials is rich, and that outer orbits can follow a wide range of evolution paths, including chaotic and centrophilic orbits \citep{merritt_book}. For simplicity and illustration purposes, we shall focus on weakly triaxial potentials and outer orbits with modest eccentricities. These lead to toroidal orbits, similar to the axisymmetric case, but with circulation of one of the symmetry axes. These approximations will allow us to analytically explore these dynamics in certain limiting cases. We leave a full exploration of the phase space to future works.

\begin{figure}[t]
    \centering
    \includegraphics[width=\linewidth]{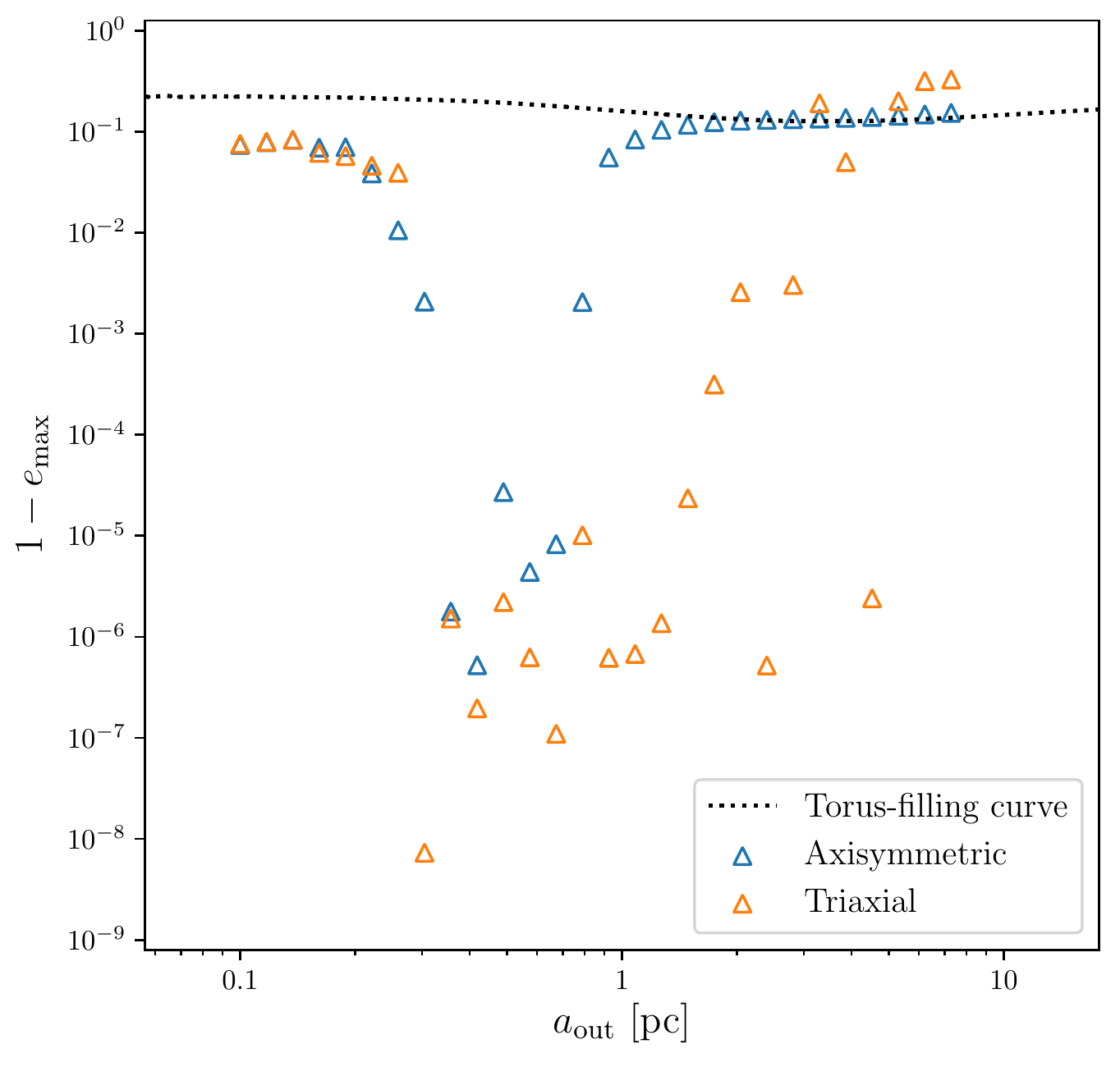}
    \caption{$1 - e_\mathrm{max}$ as a function of $a_\mathrm{out}$ for binaries in an axisymmetric and triaxial Hernquist potential. Both potentials are taken with $c = 0.7$, and the triaxial potential is taken with $b = 0.95$. The binaries are each given an outer inclination of $i_\mathrm{out} = 40^\circ$, an initial inner inclination of $i_0 = 60^\circ$, and a semi-major axis of $a_\mathrm{in} = 100 \, \mathrm{AU}$. The outer orbits are taken to be circular. The addition of triaxiality substantially widens the region of phase space where large eccentricities occur for these particular parameters.}
    \label{fig:emax_triaxial}
\end{figure}

An example of the effect of triaxiality is given in Figure \ref{fig:emax_triaxial}, which compares $1 - e_\mathrm{max}$ for two binaries in axisymmetric and triaxial potentials, respectively, each taken with $i_\mathrm{out} = 40^\circ$. Here, we notice that in the Keplerian and chaotic regimes, the behavior is similar in the axisymmetric and triaxial cases, albeit with higher maximum eccentricities achieved in the triaxial case. In the torus-filling regime, however, the addition of triaxiality results in significant increases in the maximum eccentricity for this set of parameters. Therefore, for the remainder of this section we restrict our attention to the effect of the triaxial perturbations in the torus-filling regime, where $a_\mathrm{out} \gtrsim 1 \, \mathrm{pc}$.

\subsection{Time-Averaged Equations}

\subsubsection{Potential}
We modify Equation (\ref{eq:doubly_averaged_pot}) to determine the effect of triaxiality in the torus-filling regime. In this case, the diagonal terms of the tidal tensor still vanish, however  $\langle \Phi_{xx} \rangle \neq \langle \Phi_{yy} \rangle$ (see Section \ref{sec:sim_triaxial} for a detailed example illustrating this limit). Thus, the time-averaged potential becomes
\begin{align}
\begin{split}
\label{eq:triaxial_averaged_pot}
    \langle \langle \Phi \rangle \rangle &= \frac{a_\mathrm{in}^2}{4} \sum_{i=x,y,z} \langle \Phi_{ii} \rangle \left[ 5 (\nhat_i \cdot \evec)^2 - (\nhat_i \cdot \jvec)^2 + j^2 \right] \\
    &= \frac{3 a_\mathrm{in}^2}{2} \langle \Phi_{zz} + \Phi_{yy} \rangle \big[\tfrac{1}{2} \Gamma (5 e_z^2 - j_z^2) + \tfrac{1}{2} \Pi (5 e_x^2 - j_x^2) \\ & \pushright{+ \tfrac{1}{4}e^2(1 - 5\Gamma - 2\Pi) \big]}
\end{split}
\end{align}
where
\begin{equation}
\label{eq:Gamma_Pi}
    \Gamma = \frac{\langle \Phi_{zz} - \Phi_{yy} \rangle}{3 \langle \Phi_{zz} + \Phi_{yy} \rangle} \hspace{2em} \Pi = \frac{\langle \Phi_{xx} - \Phi_{yy} \rangle}{3 \langle \Phi_{zz} + \Phi_{yy} \rangle}.
\end{equation}
Note that we slightly adjust our notation here such that $\Gamma$ is defined in terms of $\langle \Phi_{yy} \rangle$ rather than $\langle \Phi_{xx} \rangle$. We choose to do this because $\langle \Phi_{yy} \rangle$ is often larger in magnitude than $\langle \Phi_{xx} \rangle$ when setting $b < 1$ in the density profile of our nuclear star cluster. As such, $\langle \Phi_{yy} \rangle$ is more relevant for defining the secular timescale, in this case.

From the potential, we can write down a dimensionless Hamiltonian for the system given by
\begin{equation}
\label{eq:hamiltonian}
    \mathcal{H} = \frac{\langle \langle \Phi \rangle \rangle}{\Phi_0}
\end{equation}
where
\begin{equation}
\label{eq:phi_0}
    \Phi_0 = \frac{3 a_\mathrm{in}^2}{2} \langle \Phi_{zz} + \Phi_{yy} \rangle
\end{equation}
and for which the dimensionless timescale is $\tau=t/\tau_\mathrm{sec}$ with
\begin{equation}
\label{eq:secular_timescale_y}
    \tau_\mathrm{sec}^{-1} = \frac{3 a_\mathrm{in}^{3/2}}{2 \sqrt{G M_\mathrm{bin}}} \langle \Phi_{zz} + \Phi_{yy} \rangle.
\end{equation}
When writing the dimensionless Hamiltonian in terms of classical orbital elements, the canonical action-angle variables are $\{j = \sqrt{1-e^2}, \, \omega\}$ and $\{j_z = \sqrt{1 - e^2} \cos i, \, \Omega\}$, assuming nonzero inclination. In the zero-inclination case, the system has only 2 degrees of freedom, and the action-angle coordinates become $\{1 - j, \, -\varpi\}$, where $\varpi = \omega + \Omega$. We will alternate between the vectorial notation and the classical orbital elements in our following analysis. 

\begin{figure}[ht]
    \centering
    \includegraphics[width=\linewidth]{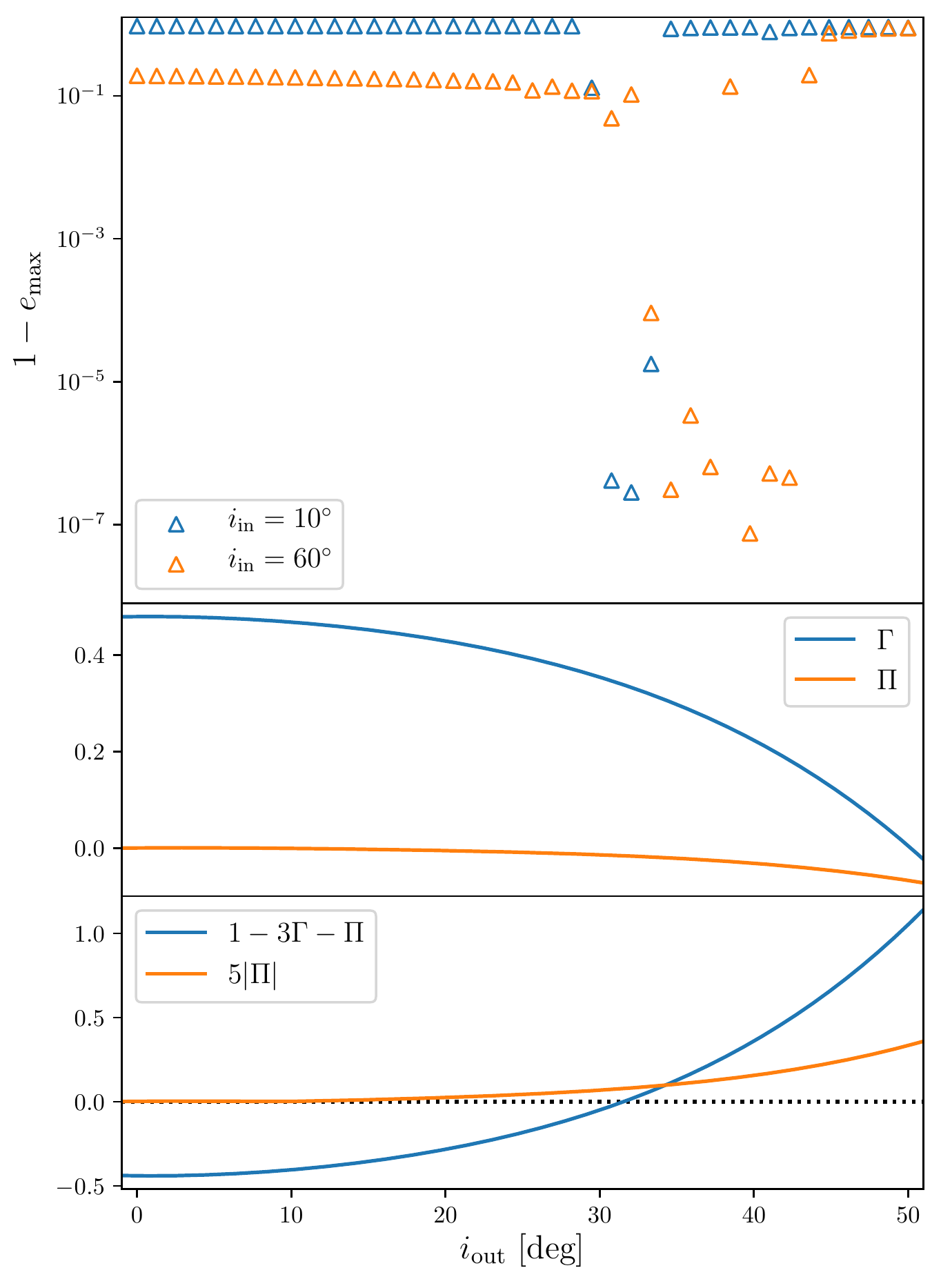}
    \caption{$1 - e_\mathrm{max}$, $\Gamma$, and $\Pi$ as a function of $i_\mathrm{out}$ for binaries in a triaxial Hernquist potential, taken with $c = 0.7$ and $b = 0.95$. The outer orbits are each taken to be circular with $a_\mathrm{out} = 1.5 \, \mathrm{pc}$, and the inner orbits are each taken with $a_\mathrm{in} = 100 \, \mathrm{AU}$, and given both $i_0 = 10^\circ$ and $i_0 = 60^\circ$. The location of the eccentricity excitations in the $i_0 = 10^\circ$ case agrees with the coplanar prediction of Equation (\ref{eq:flip_condition}). Eccentricity excitations in the $i_0 = 60^\circ$ case are driven by additional effects, such as modulation of $j_z$ (see Figure \ref{fig:evolution_triaxial}).} 
    \label{fig:emax_gamma_inc_pi}
\end{figure}

\begin{figure*}[ht]
    \centering
    \includegraphics[width=\linewidth]{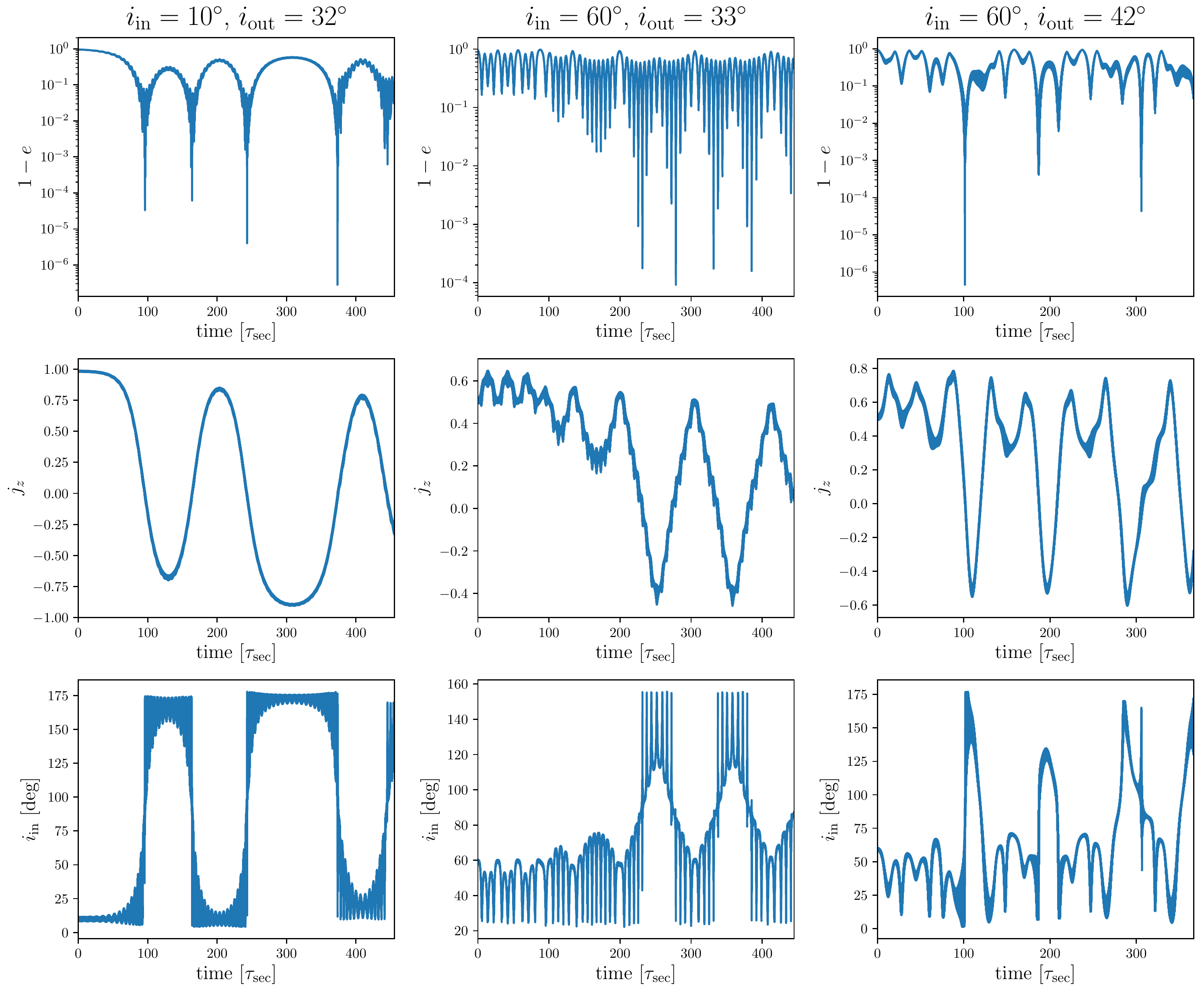}
    \caption{Evolution of binaries embedded in a triaxial Hernquist potential at $a_{\rm out}=1.5$ pc, selected from Figure \ref{fig:emax_gamma_inc_pi}. \textbf{Top row:} $1 - e$ as a function of time. \textbf{Middle row:} $j_z$ as a function of time. \textbf{Bottom row:} $i_\mathrm{in}$ as a function of time. \textbf{Left column:} Nearly coplanar binary, exhibiting extreme eccentricity excitations corresponding to retrograde flips of its orbit. \textbf{Middle column:} Binary exhibiting quadrupole-like eccentricity modulation on short timescales, which reach extreme eccentricities due to the modulation of $j_z$. \textbf{Right column:} Binary exhibiting what is likely a multitude of overlapping effects, resulting in an apparently chaotic evolution.}
    \label{fig:evolution_triaxial}
\end{figure*}

\begin{figure}[ht]
    \centering
    \includegraphics[width=\linewidth]{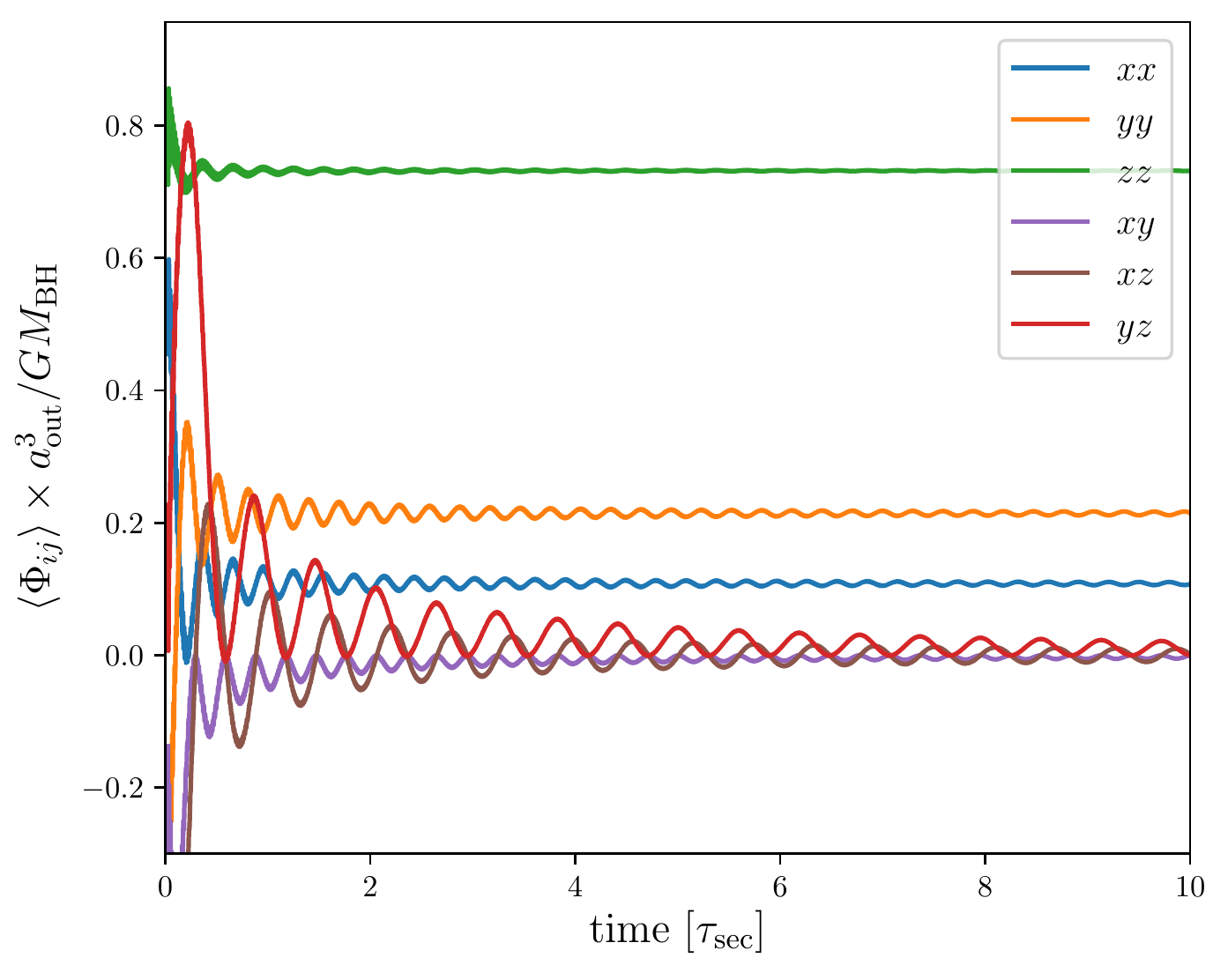}
    \caption{Averaged tidal tensor components as a function of time for the model displayed in the right column of Figure \ref{fig:evolution_triaxial}. After several secular timescales, the cross terms of the tidal tensor vanish, and the diagonal terms converge to distinct values.}
    \label{fig:triaxial_ttensor}
\end{figure}

\subsubsection{Coplanar Eccentricity Excitation}
In the limit of zero-inclination inner orbits, we have 
$j_z=\sqrt{1-e^2}$, $j_x=e_z=0$, 
and $e_x=e\cos(\varpi)$. Thus, from Equation (\ref{eq:triaxial_averaged_pot}) we can write the dimensionless Hamiltonian as
\begin{equation}
\mathcal{H}_{\rm cop}=\frac{e^2}{4}\left(1-3\Gamma\right)+
    \frac{\Pi e^2}{4}\left(5\cos^2 \varpi-1\right).
\end{equation}
Then, from Hamilton's equations,
\begin{align}
    \frac{de}{d\tau} &= -\frac{5\Pi}{2}e(1-e^2)^{1/2}\sin 2\varpi \\
    \frac{d\varpi}{d\tau} &= -\frac{(1-e^2)^{1/2}}{2}\left[1-3\Gamma+\Pi\left(5\cos^2 \varpi-1\right)\right].
\end{align}
These equations imply that the eccentricity can grow significantly if either the level of triaxiality in the potential $\Pi$ is large or $\varpi$ precesses slowly. The typical timescale for the eccentricity growth is
\begin{equation}
    \tau_{\rm sec,triaxial}=\frac{\tau_{\rm sec}}{\left|\Pi\right|}.
\end{equation}

Furthermore, we can integrate these equations of motion by writing
\begin{equation}
    \frac{de}{d\varpi}=\frac{5\Pi e \sin 2\varpi}{1-3\Gamma+\Pi\left(5\cos^2 \varpi-1\right)}
\end{equation}
and specifying an initial condition $(e_0, \, \varpi_0)$, such that
\begin{equation} \label{eq:e_varpi}
    e(\varpi)/e_0=\frac{1-3\Gamma+\Pi\left(5\cos^2 \varpi_0-1\right)}{1-3\Gamma+\Pi\left(5\cos^2 \varpi-1\right)}.
\end{equation}
Thus, it becomes clear that a necessary condition for $e \to 1$ is that the denominator vanishes or, for $\Pi<0$, that the system satisfies
\begin{equation}
\label{eq:flip_condition}
    0\leq1-3\Gamma-\Pi \leq 5|\Pi|.
\end{equation}

\subsubsection{Modulation of $j_z$}
In the general time-averaged triaxial case, $j_z$ is no longer conserved as in the torus-averaged axisymmetric case, but rather undergoes modulation. We can again use the Hamiltonian to derive this modulation as
\begin{equation}
\label{eq:djz_dtau}
    \frac{dj_z}{d \tau} = -\frac{\partial \mathcal{H}}{\partial \Omega} = \Pi \left[ 5 e_x e_y - j_x j_y \right].
\end{equation}
Thus, $j_z$ undergoes modulation with an amplitude determined by $\Pi$. This is analogous to the well-studied Lidov-Kozai mechanism, where $j_z$ can slowly change due to non-axisymmetric octupole-level perturbations \citep{katz2011,LN2011}. We will show that this modulation contributes to binaries wandering through phase portraits, and consequently encountering regimes that can lead to extreme eccentricity growth.

\subsection{Simulation Results}\label{sec:sim_triaxial}
Figure \ref{fig:emax_gamma_inc_pi} plots the maximum eccentricity as a function of $i_\mathrm{out}$ for binaries in a triaxial Hernquist potential taken at $a_\mathrm{out} = 1.5 \, \mathrm{pc}$, together with $\Gamma$ and $\Pi$. At lower inclinations, the binaries exhibit moderate maximum eccentricities, similar to the torus-averaged predictions in the axisymmetric case. At larger inclinations, the magnitude of $\Pi$ increases, and as such we observe large maximum eccentricities that do not occur in the axisymmetric case. In the case $i_\mathrm{in} = 10^\circ$, the location of these eccentricity excitations is near to the region where $0 \leq 1 - 3\Gamma - \Pi \leq 5|\Pi|$, as predicted for the coplanar case in Equation (\ref{eq:flip_condition}). In the $i_\mathrm{in} = 60^\circ$ case, large eccentricities occur up to approximately $45^\circ$. Beyond this point, we have $\Gamma < 1/5$, causing all eccentricity excitations to cease \citep{HR2019a,HR2019b}.

Full evolutions for a few selected binaries from Figure \ref{fig:emax_gamma_inc_pi} are given in Figure \ref{fig:evolution_triaxial}. Here, we can see that a variety of dynamical behaviors drive the large eccentricities observed in each binary. 

In the left column, a nearly coplanar binary exhibits a monotonically increasing eccentricity approaching unity, corresponding to a complete retrograde flip of the inner orbit. This behavior is similar to the octupole-order coplanar flipping described in \cite{li2014a}, although in this case we observe the effect at the quadrupole level.

The binary in the middle column, by contrast, shows Lidov-Kozai-like eccentricity modulation on short timescales, together with a more gradual modulation of $j_z$. As $j_z$ approaches 0, the maximum eccentricity achieved by each Lidov-Kozai cycle approaches unity. As such, this regime may be described as mimicking the torus-filling behavior on short timescales, whilst gradually wandering through phase portraits on longer timescales due to the modulation of $j_z$.

The third column displays much less clean behavior, exhibiting irregular evolution of $e$, $j_z$, and $i_\mathrm{in}$. It is possible that this binary is experiencing multiple, overlapping effects which together create an apparently chaotic evolution. For instance, at $t \approx 100 \, \tau_\mathrm{sec}$, the orbit undergoes a retrograde flip corresponding to an eccentricity spike, whereas a similar spike at $t \approx 300 \, \tau_\mathrm{sec}$ corresponds to $i_\mathrm{in}$ approaching $90^\circ$. It is also of note that this behavior occurs near to the bifurcation at $\Gamma = 1/5$ described in \cite{HR2019b}.

Finally, in Figure \ref{fig:triaxial_ttensor} we show the averaged tidal tensor components for this last example. Here, we observe that after several secular timescales all cross terms vanish, while all the diagonal terms converge to different values. Since the relevant behavior occurs over timescales $\gg\tau_{\rm sec}$, we expect that our description using a weakly distorted torus in Equation (\ref{eq:triaxial_averaged_pot}) is a good approximation to these complex dynamics.

\section{Population Synthesis} \label{sec:pop_synth}
In order to explore the overall effect of the cluster and SMBH tidal fields on mergers, we perform a population synthesis. Here, our aim is to demonstrate how the binary merger fraction varies across our fiducial models, with particular emphasis on the effect of a central SMBH and a triaxial nuclear star cluster. We also show how the merger fraction varies with distance from the Galactic center. As this work is principally focused on the dynamics at play, we leave an explicit estimate of the compact-object merger rate to future works.

\subsection{Procedure} \label{sec:sampling}
We generate a sample of binaries according to the following procedure. For the outer orbits, we begin by sampling the initial position $r_\mathrm{out}$ from a log-uniform distribution in the range $r_\mathrm{out} \in (0.1 \, \mathrm{pc}, 10 \, \mathrm{pc})$. At each such position, we sample a velocity $v$ from the distribution function $f$ of our cluster model, according to Equation (\ref{eq:velocity_pdf}). The initial position and velocity vectors $\rvec_\mathrm{out}$ and $\bm{v}$ are then each oriented by sampling azimuthal angles $\phi$ uniformly such that $\phi \sim U(0, 2\pi)$ and polar angles $\theta$ isotropically such that $\cos \theta \sim U(-1, 1)$.

For the inner orbits, we first sample $a_\mathrm{in}$ from a log-uniform distribution in the range $a_\mathrm{in} \in (10 \, \mathrm{AU}, 100 \, \mathrm{AU})$. The argument of pericenter and longitude of the ascending node are then sampled uniformly such that $\omega, \Omega \sim U(0, 2\pi)$. Initial inclinations are sampled isotropically such that $\cos i_\mathrm{in} \sim U(-1, 1)$. Finally, the initial eccentricities follow a thermal distribution in the range $e \in (0, 0.9)$, such that $e^2 \sim U(0, 0.81)$. For each of these binaries, we use the fiducial mass $M_\mathrm{bin} = 10 \, M_\odot$.

After the sampling stage, we remove binaries with unstable orbits, which would require direct N-body simulations to evolve accurately. These are given by the condition \citep{EK95,grishin2017}
\begin{equation}
\label{eq:stability}
    a_\mathrm{in}(1+e_\mathrm{in}) > 0.4 \, a_\mathrm{out} (1 - e_\mathrm{out}) \left( \frac{M_\mathrm{bin}}{3M_\mathrm{BH}} \right)^{1/3}
\end{equation}
which we evaluate at $e_\mathrm{in}=e_\mathrm{max}$.

We integrate each binary for $10 \, \tau_\mathrm{evap}$, up to a maximum of $1 \, \mathrm{Gyr}$, where $\tau_\mathrm{evap}$ is the evaporation timescale of the binary. This can be estimated as
\begin{equation}
\label{eq:tau_evap}
    \tau_\mathrm{evap} = \frac{\sqrt{3} \sigma}{32\sqrt{\pi}G a_\mathrm{in} \rho \ln \Lambda}\frac{M_\mathrm{bin}}{M}
\end{equation}
where $\sigma$ is the velocity dispersion, $\rho$ is the cluster density, $\ln \Lambda$ is the Coulomb logarithm, and $M$ is the typical mass of stars in the cluster \citep{binney+tremaine}. For the purpose of estimating the evaporation timescale, we set $\Lambda = 15$ and $M = M_\odot$. We note that although binaries are expected to evaporate after $\sim 1 \, \tau_\mathrm{evap}$, we also show results evaluated at $10 \, \tau_\mathrm{evap}$ as a proxy for systems hosting more massive binaries, which undergo more secular cycles within an evaporation time ($\tau_\mathrm{evap}/\tau_{\rm sec}\propto \sqrt{M_{\rm bin}}$) and require larger minimum pericenter distances (shorter eccentricity diffusion times) to drive gravitational wave mergers ($\tau_{\rm GW}\propto M_{\rm bin}^{-3}$).

We perform the integrations in four different potentials: a triaxial cluster, an axisymmetric cluster, a spherical cluster, and a spherical cluster without a central SMBH. Each star cluster is given a Hernquist-profile ($\gamma = 1$). Additionally, the axisymmetric and triaxial models are each initialized with $c = 0.7$, and the triaxial model is given $b = 0.95$. For the three models with an SMBH, we integrate 3,000 binaries in total. For the model without an SMBH, we integrate 10,000 binaries due to the comparatively low merger fractions and the shorter computation time for this simpler model.

\subsection{Outcomes}

We classify the outcomes of the binaries into the following categories:
\begin{itemize}
\item \textit{Gravitational wave merger}, which we define as a binary that at any point in its evolution reaches a maximum eccentricity $e_{\rm max}$ such that its inner orbit shrinks by gravitational radiation within one secular eccentricity cycle. This condition can be written as:
\begin{equation}
\label{eq:gw_condition}
    \tau_{\rm GW}<\tau_{\rm sec}\sqrt{1-e_{\rm max}^2}<1 \mbox{ Gyr} 
\end{equation}
where $\tau_{\rm GW}=(a/\dot{a})|_{\rm GW}$ is the merger timescale evaluated at the maximum eccentricity. This is given by
\begin{equation}
    \tau_{\rm GW}=\frac{3}{85}\left(\frac{a^4c^5}{G^3m_1m_2(m_1+m_2)}\right)\left(1-e_{\rm max}^2\right)^{7/2}
\end{equation}
where $m_1$ and $m_2$ are the component masses of the binary system \citep{peters64}. For our fiducial model, these are set to $m_1 = m_2 = 5 \, M_\odot$. We note that all the systems drawn have initial $\tau_{\rm GW}>1$ Gyr and would not merge if not for the effect from the cluster and/or the SMBH.

\item \textit{Tidal capture of a solar-type star}, in which a binary's pericenter shrinks to the characteristic tidal radius $r_t=R_1(m_1/m_2)^{1/3}$, such that a tidal capture is likely \citep[e.g.,][]{LO86}. Assuming a solar-type star orbiting a $9 \, M_\odot$ black hole, such that the total mass is $m_1+m_2=10 \, M_\odot$, we arrive at the following condition for a tidal capture:
\begin{equation}
    a_\mathrm{in} (1 - e_\mathrm{max}) \lesssim 2 R_\odot \approx 10^{-2} \, \mathrm{AU}
\end{equation}
which is relevant for determining formation rates of X-ray binaries \citep{Generozov2018}.
\end{itemize}
We approximate the secular timescale when computing merger fractions using Equation (\ref{eq:secular_timescale_spherical}), modifying $a_\mathrm{out} \to a_\mathrm{out} \sqrt{1 - e_\mathrm{out}^2}$ to account for elliptical orbits.

\subsection{Merger Fractions}

\begin{figure}[t]
    \centering
    \includegraphics[width=\linewidth]{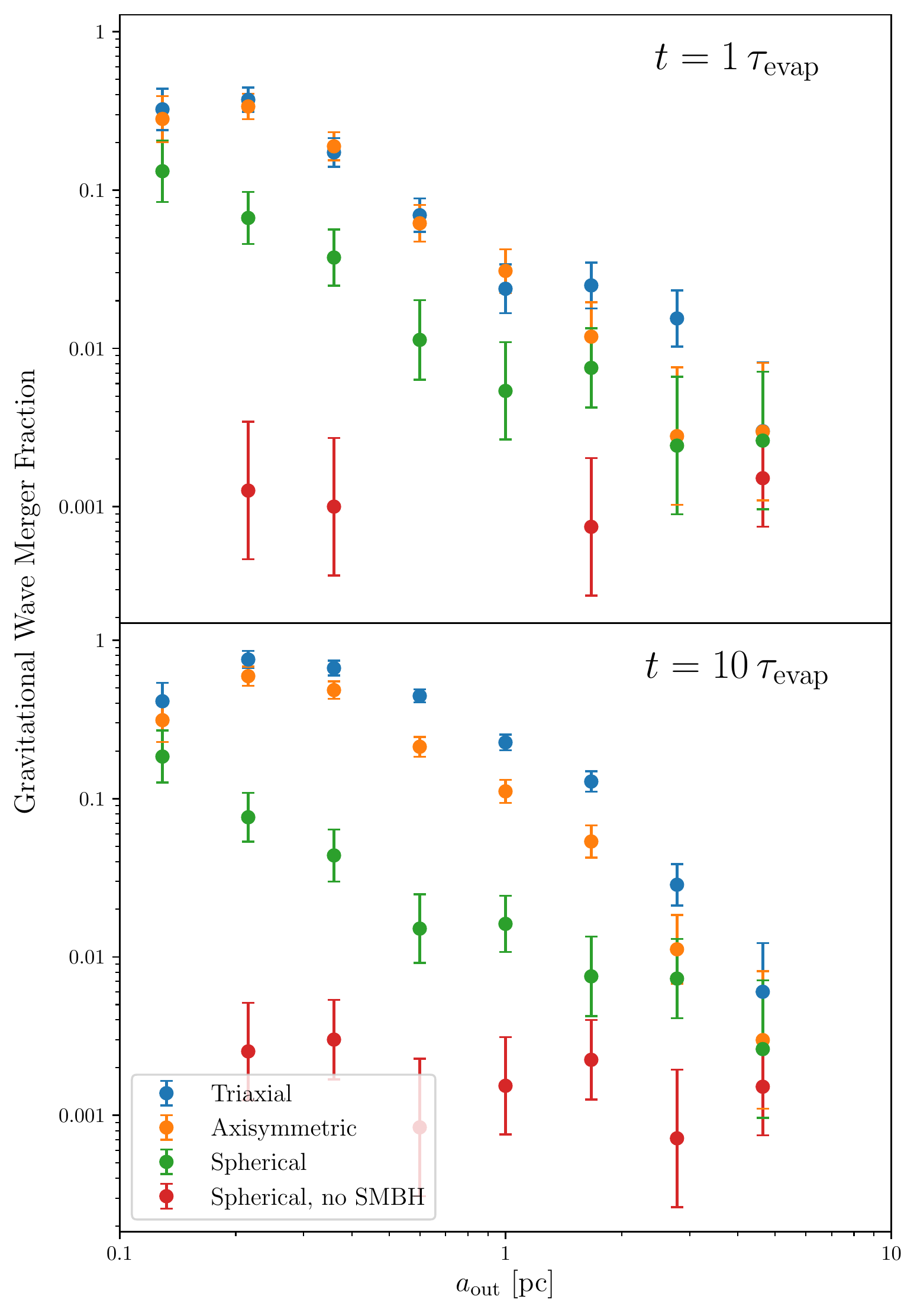}
    \caption{Gravitational wave merger fraction as a function of $a_\mathrm{out}$ for a variety of potentials. \textbf{Top panel:} Merger fractions evaluated at $t = 1 \, \tau_\mathrm{evap}$. The presence of an SMBH contributes to order-of-magnitude increases to the merger fraction in the inner parsec of spherically symmetric star clusters. The addition of triaxiality further enhances merger fractions by a factor of about 2-10. \textbf{Bottom panel:} Merger fractions evaluated at $t = 10 \, \tau_\mathrm{evap}$. Here, triaxiality enhances the merger fraction even further, with fractions approaching 70\% in the inner parsec of the Galaxy. This represents an increase by a factor of about 10-30 relative to the spherical case in this region.}
    \label{fig:gw_log_merger_frac}
\end{figure}

Figure \ref{fig:gw_log_merger_frac} shows the gravitational wave merger fraction for our population synthesis of binaries binned as a function of $a_\mathrm{out}$\footnote{We have computed $a_\mathrm{out}$ as the average of the closest and furthest approach to the SMBH throughout the integration. Given the coarse binning, this rough definition is sufficient to illustrate the sense of distance within the cluster.}.  This plot shows clearly the substantial effect of an SMBH and non-spherical nuclear star cluster on merger fractions.

In the case of a spherical star cluster without an SMBH, merger fractions are consistently low at $< 1\%$, and do not vary significantly within error as a function of $a_\mathrm{out}$. This is consistent with the results of \citet{HR2019c}. The addition of an SMBH to this spherically symmetric system causes an order-of-magnitude increase in the merger fractions, reaching approximately 10\% as one approaches $0.1 \, \mathrm{pc}$. This is largely due to the break-down of the double-averaging approximation as one approaches the SMBH; the inclusion of an SMBH decreases $\tau_\mathrm{sec}$, and as such the secular timescale approaches the dynamical timescale for binaries whose outer orbits have low pericenters. These single-averaging effects can exacerbate the maximum eccentricities achieved, due to both oscillations on the dynamical timescale and non-conservation of $j_z$. As such, merger fractions become much larger as $a_\mathrm{out}$ decreases.

In a triaxial star cluster, merger fractions increase even further, approaching unity near $0.2 \, \mathrm{pc}$ from the SMBH after $10 \, \tau_\mathrm{evap}$. As discussed in Sections \ref{sec:non-torus-filling} and \ref{sec:triaxial}, there are two significant contributing factors to the extreme merger efficiency observed here. First, in the region $a_\mathrm{out} \lesssim 1 \, \mathrm{pc}$, mergers are driven primarily by secular chaos in the non-torus-filling regime, where the secular timescale is similar to the nodal precession timescale of the binaries. This process drives the largest merger fractions observed near 0.2-0.3 pc. For binaries with $a_\mathrm{out} \gtrsim 1 \, \mathrm{pc}$, mergers are instead primarily driven by triaxial effects, including coplanar eccentricity excitation and non-conservation of $j_z$. This can be seen by comparing the triaxial and axisymmetric cases: near the chaotic regime, merger fractions are similar for both models, whereas the triaxial case exhibits notably larger merger efficiencies at $a_\mathrm{out} \gtrsim 0.5 \, \mathrm{pc}$. The combined effect is a consistently large merger fraction that is an order of magnitude larger than the equivalent spherical model in most regions. At very small and very large values of $a_\mathrm{out}$, the merger fractions begin to become similar to the spherical case, as the SMBH becomes dominant and the cluster appears increasingly like a point mass, respectively.

Comparing the $t = 1 \, \tau_\mathrm{evap}$ and $t = 10 \, \tau_\mathrm{evap}$ panels in this plot reveals the substantial effect of the integration time. Indeed, we see that most binaries which merge in the triaxial and axisymmetric cases do so after $t = 1 \, \tau_\mathrm{evap}$, and that the distinctions between the triaxial and axisymmetric cases do not become apparent until the later time. As such, it will important for future works to examine how the collisional dynamics affect these considerations.

\begin{figure}[t]
    \centering
    \includegraphics[width=\linewidth]{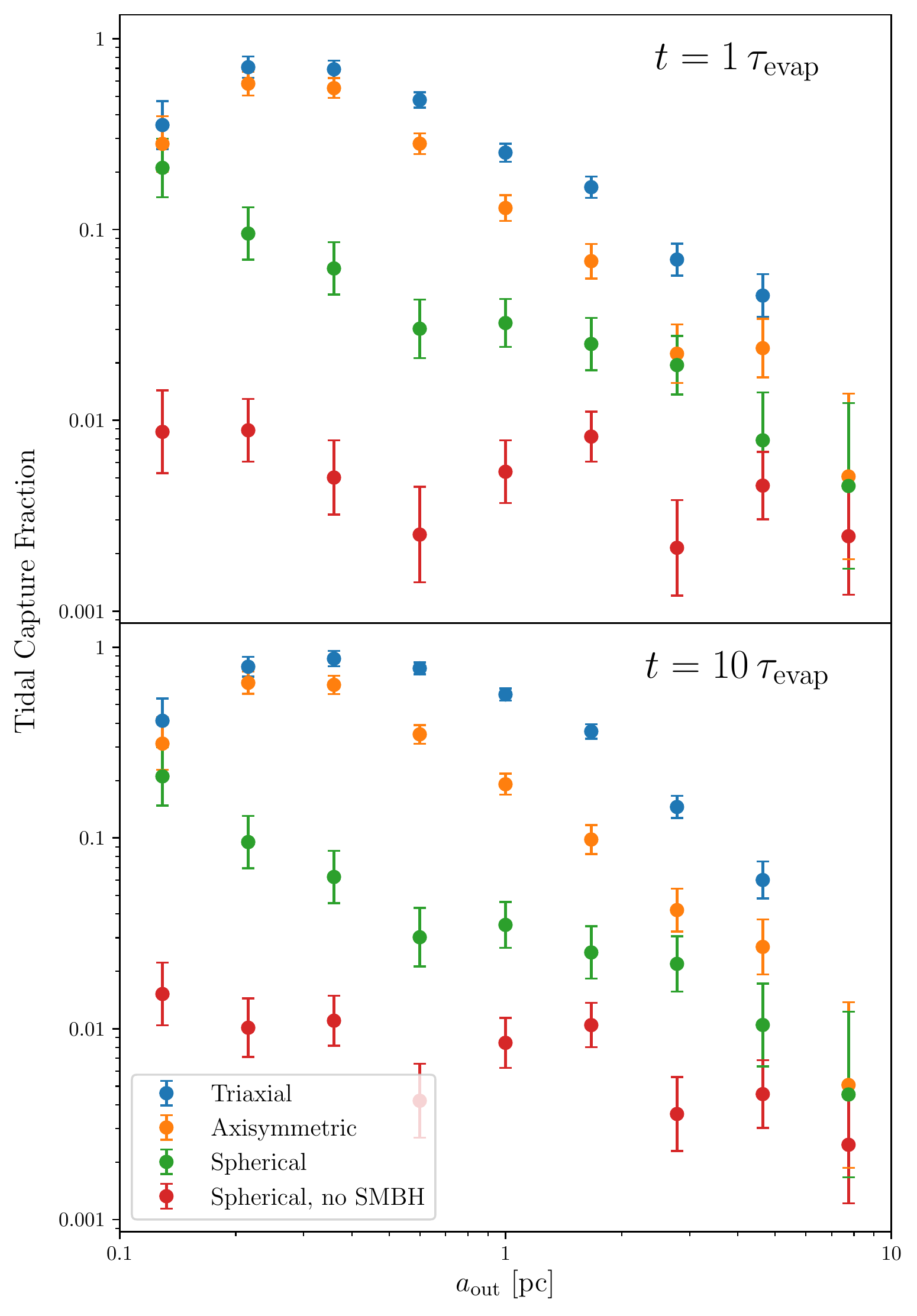}
    \caption{Tidal capture fraction as a function of $a_\mathrm{out}$ for a variety of potentials. \textbf{Top panel:} Capture fractions evaluated at $t = 1 \, \tau_\mathrm{evap}$. Fractions are again significantly enhanced by the presence of an SMBH and triaxial nuclear star cluster. Because the tidal capture condition is somewhat more permissive than the gravitational wave merger condition, these fractions are higher than those in Figure \ref{fig:gw_log_merger_frac}, even after only $1 \, \tau_\mathrm{evap}$ of integration time. \textbf{Bottom panel:} Capture fractions evaluated at $t = 10 \, \tau_\mathrm{evap}$. After the full integration time, the capture fractions reach as high as 90\% near $0.3 \, \mathrm{pc}$ in the triaxial case.}
    \label{fig:capture_log_merger_frac}
\end{figure}

Figure \ref{fig:capture_log_merger_frac} shows the tidal capture fraction as a function of $a_\mathrm{out}$ for our four fiducial potentials. In comparison with Figure \ref{fig:gw_log_merger_frac}, the overall trends are similar, with order-of-magnitude increases in the tidal capture fractions with the addition of an SMBH and triaxial nuclear star cluster. Capture fractions are higher overall than with the gravitational wave merger condition, since the tidal capture condition tends to require somewhat lower maximum eccentricities. As such, in the triaxial case the capture fraction reaches as high as 90\% in the chaotic regime. Additionally, the capture fractions do not drop off as quickly at large $a_\mathrm{out}$ as they do with the gravitational wave condition. This is likely because binaries at these large distances have much longer secular timescales, and as such have fewer opportunities during the integration time to reach the large eccentricities required for a gravitational wave merger. That is to say, there are a number of binaries which have enough time to satisfy the more permissive tidal capture condition, but would require a longer integration time to satisfy the gravitational wave condition.

Note that Figures \ref{fig:gw_log_merger_frac} and \ref{fig:capture_log_merger_frac} are missing data points in bins where no binaries in our population synthesis merged, particularly near the left and right edges of the plots. This is largely due to the sampling procedure, which produces binaries with $a_\mathrm{out}$ near $0.1 \, \mathrm{pc}$ and $10 \, \mathrm{pc}$ less frequently than it produces binaries with $a_\mathrm{out}$ closer to the middle of the plot range. The number of binaries toward the edges is also influenced by the stability criterion (\ref{eq:stability}), which reduces the number of binaries with low $a_\mathrm{out}$.

\section{Discussion}\label{sec:discussion}
In this work, we studied the secular dynamics of stellar binaries in the Galactic center, accounting for both the potential from a central SMBH and a triaxial nuclear star cluster. Our main result is that even modest levels of triaxiality (with axis ratios of $0.7$ and $0.95$) can dramatically enhance the compact-object merger fractions in the center of the Galaxy, by a factor of $\sim10-30$ relative to a spherical cluster. Moreover, these merger fractions reach near-unity values at $\sim0.2-0.4$ pc from the Galactic center, with fractions remaining above $\sim 10\%$ in the central $\sim2$ pc.

These results demonstrate that compact-object mergers in galactic nuclei driven by secular dynamics are not confined to the innermost $0.1 \, \mathrm{pc}$ of the cluster \citep[e.g.,][]{hoang2018}, but rather could reach up to the effective radii of the cluster. In turn, this implies enhanced rates of compact-object mergers by gravitational radiation compared to previous studies, and the possibility of forming the X-ray binaries by tidal captures in the inner parsec of the cluster \citep{Generozov2018}. Further work including star formation and evolution is required to quantify these formation rates.

We have also developed a code\footnote{\url{https://github.com/mwbub/binary-evolution}} to evolve Galactic-center binaries in arbitrary orbits and nearly arbitrary potentials, which is a hybrid of \texttt{galpy} \citep{bovy2015} and our singly-averaged equations of motion (see Equations \ref{eq:djdt} and \ref{eq:dedt}). This is similar to the singly-averaged code implemented by \cite{HR2019a,HR2019b}, but unlike their work using orbital elements, we have expressed the equations of motion using the vectorial formalism, which has the advantage of being more compact and non-divergent\footnote{The classical Delaunay orbital elements are ill-defined for polar orbits, and as such the equations of motion diverge.}. The extensive library of potentials provided by \texttt{galpy} allows us to evolve binaries in a wide variety of environments. Although we have only applied this code to the Galactic center, it can in principle be used to study binaries in other environments (e.g., wide binaries in the Galactic field).

Other significant and more specific results from this work include:
\begin{itemize}
    \item We find that triaxial clusters lead to a new dynamical behavior in which $j_z$---the binary's angular momentum along the cluster's $z$-axis---is slowly modulated, leading to near-unity eccentricities for a wide range of orbital parameters. We understand this behavior in two limits displayed in Figure \ref{fig:evolution_triaxial}. First, for highly inclined binaries (middle column), the slow modulation enhances the eccentricity oscillations driven by the axisymmetric part of the potential, similar to the well-studied octupole-level modulations of the Lidov-Kozai mechanism in three-body systems \citep{naoz2016}. Second, for low-inclination binaries (left column), binaries are slowly torqued by the triaxial part of the potential, leading to regular eccentricity growth coupled with orbit flipping, analogous to its counterpart in coplanar three-body systems \citep{li2014a}.
    
    \item We confirm the existence of a \emph{chaotic regime} where a large fraction of binaries are excited to extreme eccentricities, as previously discovered by \citet{petrovich2017} using doubly-averaged equations in axisymmetric clusters. Here, secular chaos arises when the nodal precession timescale of the binary's outer orbit about the SMBH approaches the secular timescale. We show that an equivalent interpretation of this regime within the framework of \citet{HR2019a, HR2019b} is that chaos occurs when the binary evolution is neither approximated by a three-body system nor by the tidal field averaged over an axisymmetric torus. Here, the presence of an SMBH significantly slows the nodal precession rate, such that orbits do not densely fill an axisymmetric torus within a secular timescale. This can be seen in the middle column of Figure \ref{fig:e_xy_tt}. Furthermore, we find that triaxial clusters expand the available phase space for secular chaos compared to axisymmetric clusters, allowing for chaos to occur at larger distances from the SMBH (see Figure \ref{fig:emax_triaxial}). 

    \item  We find that for spherical clusters the presence of an SMBH greatly increases the fraction of mergers in the inner parsec of the cluster. This is related to at least two separate effects. First, the SMBH allows for a larger concentration of mass inside a binary's orbit. In turn, this keeps the value of $\Gamma$ as defined in Equation (\ref{eq:Gamma}) above the critical value of 1/5, below which eccentricity excitations do not occur \citep{HR2019a, HR2019b}. Second, the presence of an SMBH leads to a break-down of the double-averaging approximation for a wide range of binaries (see Figure \ref{fig:regimes}), significantly enhancing the rate of mergers near the SMBH (see Figure \ref{fig:gw_log_merger_frac}, spherical case). This latter result is consistent with previous N-body experiments \citep[e.g.,][]{antonini2012, fragione_antonini2019}.
\end{itemize}

Overall, our results show that the level of triaxiality of nuclear star clusters plays a major role at determining the merger frequencies of binaries, revealing in particular a link between the morphology of the centers of galaxies and enhanced rates of gravitational wave mergers.

\section*{Acknowledgments}
We thank Chris Hamilton for providing helpful comments on the manuscript. We are grateful to Roman Rafikov, Diego Munoz, Katie Breivik, Almog Yalinewich, Norm Murray, Hagai Perets, and John Dubinski for stimulating useful discussion. M.~Bub acknowledges the support of the Natural Sciences and Engineering Research Council of Canada (NSERC; funding reference number USRA-540486-2019). C.~Petrovich acknowledges support from the Gruber Foundation Fellowship and Jeffrey L. Bishop Fellowship at CITA, and the Bart J. Bok fellowship at Steward Observatory. This research made use of \texttt{astropy}, a community-developed core Python package for astronomy \citep{astropy2013, astropy2018}, and \texttt{galpy} \citep{bovy2015}.

\bibliography{refs}

\appendix
\section{Equations of motion} \label{ap:equations_of_motion}
From the tidal approximation given in Equation (\ref{eq:taylor}), the potential averaged over the period of the inner binary is given by
\begin{equation}
\label{eq:averaged_pot_base}
    \langle \Phi \rangle = \frac{1}{2} \sum_{i,j=x,y,z} \Phi_{ij}(\rvec) \langle x_i x_j \rangle.
\end{equation}
Writing $x_i = \nhat_i \cdot \rvec_\mathrm{in}$, where $\rvec_\mathrm{in}$ is the displacement vector of the inner binary, we may find an explicit form for this potential by calculating $\langle x_i x_j \rangle = \langle (\nhat_i \cdot \rvec_\mathrm{in})(\nhat_j \cdot \rvec_\mathrm{in}) \rangle$. To do so, we write $\rvec_\mathrm{in} = r \cos \phi \, \ehat + r \sin \phi \, \qhat$, where $\phi$ is the true anomaly and $\qhat = \jhat \times \ehat$. We then have that
\begin{equation}
\begin{multlined}
\label{eq:xixj}
    (\nhat_i \cdot \rvec_\mathrm{in})(\nhat_j \cdot \rvec_\mathrm{in}) = r^2 \big[ \cos^2 \phi \, (\nhat_i \cdot \ehat)(\nhat_j \cdot \ehat) + \cos \phi \sin \phi \, (\nhat_i \cdot \ehat)(\nhat_j \cdot \qhat) \\ + \cos \phi \sin \phi \, (\nhat_i \cdot \qhat)(\nhat_j \cdot \ehat) + \sin^2 \phi \, (\nhat_i \cdot \qhat)(\nhat_j \cdot \qhat) \big].
\end{multlined}
\end{equation}
To average Equation (\ref{eq:xixj}), therefore, we compute $\langle r^2 \cos^2 \phi \rangle$, $\langle r^2 \sin^2 \phi \rangle$, and $\langle r^2 \cos \phi \sin \phi \rangle$. The average of some arbitrary function $f(\rvec_\mathrm{in})$ over the orbit of the inner binary is given by
\begin{equation}
\label{eq:averaged_func}
    \langle f(\rvec_\mathrm{in}) \rangle = \frac{(1 - e^2)^{3/2}}{2 \pi} \int_0^{2 \pi} \frac{d \phi}{(1 + e \cos \phi)^2} f(r, \phi)
\end{equation}
\citep{tremaine2014}. Using this expression, as well as the fact that $r = a (1 - e^2) / (1 + e \cos \phi)$, we find that
\begin{align}
\begin{split}
\label{eq:averaged_terms}
    \langle r^2 \cos^2 \phi \rangle &= \frac{a^2}{2} (1 + 4 e^2) \\
    \langle r^2 \sin^2 \phi \rangle &= \frac{a^2}{2} (1 - e^2) \\
    \langle r^2 \cos \phi \sin \phi \rangle &= 0.
\end{split}
\end{align}
Thus, we have that
\begin{equation}
\label{eq:xixj_averaged}
    \langle (\nhat_i \cdot \rvec_\mathrm{in})(\nhat_j \cdot \rvec_\mathrm{in}) \rangle = \frac{a^2}{2} \left[ (1 + 4e^2)(\nhat_i \cdot \ehat)(\nhat_j \cdot \ehat) + (1 - e^2)(\nhat_i \cdot \qhat)(\nhat_j \cdot \qhat) \right].
\end{equation}
We wish to eliminate $\qvec$ from this expression in favor of $\jvec$. To do so, we can expand the simple product of two triple products as
\begin{align}
\begin{split}
\label{eq:triple_products}
    (\nhat_i \cdot \qhat)(\nhat_j \cdot \qhat) &= \left( (\jhat \times \ehat) \cdot \nhat_i \right) \left( (\jhat \times \ehat) \cdot \nhat_j \right) \\
    &= \begin{vmatrix}
        \jhat \cdot \jhat & \jhat \cdot \ehat & \jhat \cdot \nhat_j \\
        \ehat \cdot \jhat & \ehat \cdot \ehat & \ehat \cdot \nhat_j \\
        \nhat_i \cdot \jhat & \nhat_i \cdot \ehat & \nhat_i \cdot \nhat_j
    \end{vmatrix} \\
    &= \delta_{ij} - (\nhat_i \cdot \ehat)(\nhat_j \cdot \ehat) - (\nhat_i \cdot \jhat)(\nhat_j \cdot \jhat)
\end{split}
\end{align}
where $\delta_{ij}$ is the Kronecker delta. Substituting this into Equation (\ref{eq:averaged_pot_base}) gives
\begin{equation}
\label{averaged_pot_appendix}
    \langle \Phi \rangle = \frac{a^2}{4} \sum_{i,j = x,y,z} \Phi_{ij} (\rvec) \left[ 5(\nhat_i \cdot \evec)(\nhat_j \cdot \evec) - (\nhat_i \cdot \jvec)(\nhat_j \cdot \jvec) + j^2 \delta_{ij} \right].
\end{equation}

From the singly-averaged potential, the secular evolution is given by the Milankovitch's equations of motion 
\begin{align}
    \frac{d \jvec}{dt} &= -\frac{1}{\sqrt{GM_\mathrm{bin}a}} (\jvec \times \nabla_{\jvec} \langle \Phi \rangle + \evec \times \nabla_{\evec} \langle \Phi \rangle ) \\
    \frac{d \evec}{dt} &= -\frac{1}{\sqrt{GM_\mathrm{bin}a}} (\jvec \times \nabla_{\evec} \langle \Phi \rangle + \evec \times \nabla_{\jvec} \langle \Phi \rangle ) 
\end{align}
where $\nabla_{\jvec} \equiv (\partial / \partial j_x, \, \partial / \partial j_y, \partial / \partial j_z)$ and $\nabla_{\evec} \equiv (\partial / \partial e_x, \, \partial / \partial e_y, \partial / \partial e_z)$ \citep[e.g., ][]{tremaine2009}. Performing this calculation gives the equations of motion as
\begin{align}\label{eq:dj_dt_appendix}
    \frac{d \jvec}{dt} &= \frac{a^{3/2}}{2 \sqrt{G M_\mathrm{bin}}} \sum_{i,j=x,y,z} \Phi_{ij}(\rvec) [(\nhat_j \cdot \jvec)(\jvec \times \nhat_i) - 5(\nhat_j \cdot \evec)(\evec \times \nhat_i)] \\
    \label{eq:de_dt_appendix}
    \frac{d \evec}{dt} &= \frac{a^{3/2}}{2 \sqrt{G M_\mathrm{bin}}} \sum_{i,j=x,y,z} \Phi_{ij}(\rvec) [(\nhat_j \cdot \jvec)(\evec \times \nhat_i) - 5(\nhat_j \cdot \evec)(\jvec \times \nhat_i) + \delta_{ij} (\jvec \times \evec)].
\end{align}

\subsection{Keplerian Potential}

For the specific case where the potential $\Phi$ is only due to the central massive black hole, Equations (\ref{eq:dj_dt_appendix}) and (\ref{eq:de_dt_appendix}) reduce to
\begin{align}
    \frac{d\jvec}{dt} &= \tau_\mathrm{bin}^{-1} \left[ 5(\rhat \cdot \evec)(\evec \times \rhat) -(\rhat \cdot \jvec)(\jvec \times \rhat) \right] \label{eq:djdt_bh} \\
    \frac{d\evec}{dt} &= \tau_\mathrm{bin}^{-1} \left[ 5(\rhat \cdot \evec)(\jvec \times \rhat) - (\rhat \cdot \jvec)(\evec \times \rhat) - 2 (\jvec \times \evec)  \right] \label{eq:dedt_bh}
\end{align}
where
\begin{equation}
    \tau_\mathrm{bin} = \frac{M_\mathrm{bin}}{M_\mathrm{BH}} \frac{R^3}{a^3} \frac{P}{3\pi} \label{eq:tau}
\end{equation}
which is consistent with previous results \citep[e.g.,][]{LiuLai2018}. We checked that our code with a Keplerian potential from \texttt{galpy} gives the same results as those using the analytic equations above.

\section{Potential for a slightly flattened cluster mass distribution: explicit expression for a Hernquist potential} \label{ap:flattened_hernquist}
\label{sec:phi_flat}
 Let us consider the axisymmetric density profile
\begin{equation}
\label{eq:gamma_family_ap}
    \rho(x, y, z) = \frac{(3 - \gamma) M_\mathrm{cl}}{4 \pi  c} \frac{s}{m^\gamma (m + s)^{4 - \gamma}}
\end{equation}
and express the elliptical variable $m$ as $m^2=(x^2+y^2)+z^2/c^2=r^2+\epsilon_z r^2\cos ^2\theta$ with $\epsilon_z\equiv (1-c^2)/c^2$. We can expand this profile to first order in $\epsilon_z$ and
conveniently write it in terms of the second-order Legendre polynomial, $P_2(x)=\tfrac{1}{2}(3x^2-1)$, as
\begin{equation}
    \rho(r,\theta) = \frac{(3 - \gamma) M_\mathrm{cl}}{4 \pi  c} \frac{s}{r^\gamma (r + s)^{4 - \gamma}}\left[1-\frac{\epsilon_z}{3} \left(\gamma+(4-\gamma)\frac{r}{r+s}\right)\left(P_2\left(\cos \theta\right) +\tfrac{1}{2}\right) \right]
  \end{equation}  
  where positive densities are defined everywhere for $\epsilon_z<1/2$ ($c\gtrsim0.82$).
The associated potential can be obtained from Poisson's equation separating the solutions for $P_0\left(\cos \theta\right)$ and $P_2\left(\cos \theta\right)$. The solution for general $\gamma$ is given in terms hypergeometric functions and is not particularly useful to provide with simple analytical estimates. Instead, we provide the solution for the Hernquist potential ($\gamma=1$), which results in
\begin{equation}
\label{eq:hernquist_flat}
    \Phi_\mathrm{cl}(r,\theta)=-\frac{GM_\mathrm{cl}}{c(r+s)}\left[1-\frac{\epsilon_z}{6} \frac{(2s+3r)}{(s+r)} -\frac{\epsilon_z}{3(r+s)}
    \left(1 + \frac{9s}{2r}+\frac{3s^2}{r^2}
    -\frac{3s(s+r)^2}{r^3}\log(1+r/s)\right)P_2\left(\cos \theta\right)
    \right].
  \end{equation}  
It will become convenient to express this potential inside the sphere of influence of the black hole, so we expand it at $r\ll s$ to get
\begin{equation}
    \Phi_\mathrm{cl}(r,\theta)\approx -\frac{GM_\mathrm{cl}}{c(r+s)}\left\{1-\frac{\epsilon_z}{6} \frac{(2s+3r)}{s} -\frac{\epsilon_z }{12}\frac{r}{s}P_2\left(\cos \theta\right)
    + O\left[\epsilon_z (r/s)^2\right]
    \right\}.
\end{equation}

\section{Velocity dispersion and distribution function for a $\gamma$-family potential with a central black hole} \label{ap:df}

For the purpose of estimating the typical velocity dispersion in our cluster model, we assume that the system is spherical and isotropic. Thus, from the Jeans equations \citep{binney+tremaine}, we have that
\begin{equation}
\label{eq:sigma_integral}
    \sigma^2(r) = \frac{1}{\rho(r)} \int_r^\infty \frac{M_\mathrm{encl}(r') \rho(r')}{r'^2} \, dr'
\end{equation}
where $M_\mathrm{encl}(r)$ is the enclosed mass. This integral can be computed analytically for various values of $\gamma$ \citep[e.g.,][]{tremaine1994, baes2005}. In particular, for the Hernquist profile, we have that
\begin{equation}
\label{eq:sigma_hernquist}
    \sigma^2(r) = \frac{G M_\mathrm{BH}}{s} F_1(r/s) + \frac{G M_\mathrm{cl}}{s} F_2(r/s)
\end{equation}
where
\begin{align}
    F_1(x) &= 6 x (1 + x)^3 \ln(1 + 1/x) + \frac{1}{2x} - \tfrac{3}{2} - 11x - 15x^2 - 6x^3 \\
    F_2(x) &= x (1 + x)^3 \left[ \ln(1 + 1/x) - \tfrac{25}{12} \right] + 4 x^2 (1 + x)^2 - 3 x^3 (1 + x) + \tfrac{4}{3} x^4 - \frac{x^5}{4(1+x)}
\end{align}
\citep{tremaine1994}.

The distribution function $f$ for the combined SMBH plus spherical $\gamma$-family potential can be computed numerically via Eddington's formula:
\begin{equation}
\label{eq:eddington}
    f(\mathcal{E}) = \frac{1}{\sqrt{8}\pi^2} \left[ \int_0^\mathcal{E} \frac{d \Psi}{\sqrt{\mathcal{E} - \Psi}} \frac{d^2 \nu}{d\Psi^2} + \frac{1}{\sqrt{\mathcal{E}}} \frac{d \nu}{d \Psi} \bigg|_{\Psi = 0} \right]
\end{equation}
where $\Psi \equiv -\Phi$ is the relative potential, $\mathcal{E} \equiv \Psi - \tfrac{1}{2} v^2$ is the relative energy, and $\nu(r) \equiv \rho(r) / M_\mathrm{cl}$ is the spatial probability density \citep{binney+tremaine}. Note that the second term in this expression vanishes for the $\gamma$-family. Also note that the numerical evaluation of this integral can be simplified by transforming the problem to be expressed entirely in terms of the cluster potential, as described in \cite{baes2005}. We will not reproduce these transformations here, but rather refer the reader to the aforementioned work.

The velocity probability density at position $r$ can be obtained from the distribution function as
\begin{equation}
\label{eq:velocity_pdf}
    P(v, r) = \frac{f(r, v)}{\nu(r)}
\end{equation}
\citep{binney+tremaine}. We use this expression to sample velocities for the population synthesis (Section \ref{sec:pop_synth}).

\end{document}